\UseRawInputEncoding
\documentclass[aps, prx, amsmath, amssymb, reprint]{revtex4-2}
\usepackage{lipsum} 
\usepackage{graphicx}
\usepackage{dcolumn}
\usepackage{color}
\usepackage[normalem]{ulem} %strike out tool using \sout{...}
\usepackage{cancel}
\usepackage{amsmath,amsfonts,mathtools}
\begin{document}
\title{Free-space subterahertz field-polarization controlled by waveguide-mode-selection}
\author{Marc Westig}
\email{mpwestig@gmail.com}
\affiliation{Kavli Institute of NanoScience, Delft University of Technology, 
Lorentzweg 1, 2628 CJ Delft, The Netherlands}
\author{Holger Thierschmann}
\affiliation{Kavli Institute of NanoScience, Delft University of Technology, 
Lorentzweg 1, 2628 CJ Delft, The Netherlands}
\author{Allard Katan}
\affiliation{Kavli Institute of NanoScience, Delft University of Technology, 
Lorentzweg 1, 2628 CJ Delft, The Netherlands}
\author{Matvey Finkel}
\affiliation{Kavli Institute of NanoScience, Delft University of Technology, 
Lorentzweg 1, 2628 CJ Delft, The Netherlands}
\author{Teun M. Klapwijk}
\affiliation{Kavli Institute of NanoScience, Delft University of Technology, 
Lorentzweg 1, 2628 CJ Delft, The Netherlands}
\begin{abstract}
We study experimentally the free-space electro-magnetic field emitted 
from a multimode rectangular waveguide equipped with a diagonal-horn antenna. 
Using the frequency range of 215-580 GHz, a photo-mixer is used to launch 
a free-space circularly-polarized electro-magnetic field, exciting multiple 
modes at the input of the rectangular waveguide via an input diagonal-horn antenna. 
A second photo-mixer is used, together with a silicon mirror 
Fresnel scatterer, to act as a polarization-sensitive coherent detector to 
characterize the emitted field. We find that the radiated field, excited by 
the fundamental waveguide mode, is characterized by a linear polarization. 
In addition, we find that the polarization of the radiated field rotates by 
$45^\circ$ if selectively 
exciting higher-order modes in the waveguide.
Despite the higher-order modes, the radiated field 
appears to maintain a predominant Gaussian beam character, since an 
unidirectional coupling to a detector was possible, whereas the 
unidirectionality is independent of the frequency. 
We discuss a possible application of this finding.
\end{abstract}
\maketitle 
\section{\label{sec:01}Introduction}
Terahertz (THz) waves lie in the frequency 
range 100~GHz - 30~THz between microwaves and visible 
light and have properties in common with both frequency 
domains. For instance, THz waves show partial absorption 
or reflection from objects due to the rich THz 
excitation spectrum of matter and provide
a high-enough spatial resolution for imaging and detector applications 
because of their sub-mm wavelength. The progress in THz 
devices and measurement techniques 
\cite{natphotoneditorial2013, mittleman2013, horiuchi2013}, 
therefore, has consequently led to the need for an improved 
understanding of light properties at the intersection of free space 
optics and waveguide circuit technologies. Latter circuit technologies 
comprise planar circuits and three-dimensional conductors such 
as hollow waveguides, commonly used to controllably radiate and detect
THz waves \cite{dhillon2017}. 

In the realm of THz research, circuit quantum 
electrodynamics (cQED) has recently gained attention, 
whereas cQED commonly employs much lower frequencies 
up to about 10 GHz. In view of the work of 
\citet{wallraff2004}, which established microwave photons 
confined to a qubit circuit at milli-Kelvin temperatures 
in inter-fridge quantum experiments, THz waves with their ability to 
couple efficiently to free space would provide cQED with 
a paradigm change. The idea has recently been explored in 
the work of \citet{sanz2018_A} which proposes to 
extend cQED concepts to open-air microwave quantum 
communication, quantum illumination and quantum sensing.
This could show routes to enable transfer of quantum 
states between fridges, but seems to require antennas connecting 
in a scalable manner the respective experiments via free-space.
Accordingly, the technical concept of emission and reception of 
microwave and THz quantum signals by means of antennas 
shall be included into the cQED framework. Moreover, to push the field 
forward, a major challenge consists in further improving and 
exploring methods, compatible with cQED, to emit,
receive and analyze free space signals using antennas and hollow-waveguides.

Recently, we 
have demonstrated how THz photo-mixers \cite{deninger2013} 
can be used to probe and analyze a single-mode THz 
signal with a linear polarization, transmitted through a 
waveguide \cite{westig2020}. This has enabled us to 
study the waveguide from the perspective of a 
single-mode communication channel. 
In our previous work, we have further studied 
the suitability of waveguides to pick up non-classical signals generated by 
cQED devices \cite{rolland2019, gramich2013, dambach2015, grimm2019, 
leppakangas2015, leppakangas2016, westig2017, leppakangas2013, 
leppakangas2014, armour2015, trif2015} and to radiate 
the quantum field with a linear polarization into free space 
via a diagonal-horn antenna. In this regard, cQED devices have been 
explored as sources for the waveguide field. In particular, cQED 
devices with their large flexibility of generating various electro-magnetic fields 
characterized by quantum or classical photon statistics, could
provide essentially any desired electro-magnetic field for a large class
of applications. Independent from the technological aspects, 
including the THz domain into quantum experiments at 
milli-Kelvin temperatures would allow to conduct experiments 
in the deep quantum regime with effectively zero thermal photon population.
This may be achieved pursuing two different strategies. First, 
THz radiation can be generated by Josephson junctions, like described in 
\cite{westig2017}. However, as already described in \cite{westig2017}, 
in order to achieve this, a superconducting material with an energy gap $\Delta$ is 
needed for the Josephson junction electrodes, 
such that $2\Delta/h$ reaches the desired frequency, where $h$ is the Planck 
constant. If the material niobium is employed to fabricate the electrodes of the Josephson junction, 
the maximum frequency which may be generated can reach 
700~GHz. Other superconducting materials such as NbN or NbTiN 
with a higher energy gap than niobium have to replace 
at least one niobium electrode material of the Josephson junction in 
order to push this frequency above 700~GHz. One challenge in this more advanced 
fabrication is the material composition of NbN or NbTiN. In particular it seems advantageous 
to replace the barrier material, changing the commonly employed aluminum oxide barrier 
used frequently in cQED, to the material AlN which provides high-quality NbN 
or NbTiN junction technologies with low leakage currents and if desired also 
with high current densities \cite{Zijlstra2007, Kawamura1999, Nakamura2011}. 
On the other hand, among other 
nitride based superconductors, NbN shows a high kinetic inductance, which has 
recently enabled four-wave mixing around 100~GHz \cite{Anferov2020}. In the 
aforementioned experiments, however, the radiation may be generated within a 
cryogenic environment, i.e.~these are intra-fridge experiments. Furthermore, 
in these intra-fridge experiments, in order to reach the quantum regime, 
$h\nu \gg k_{B}T$, with $\nu$ the frequency, $T$ the temperature of the cryogenic 
environment and $k_{B}$ the Boltzmann constant.

The second strategy, compatible with the techniques presented 
in our work, is to generate the THz radiation \emph{outside} the cryogenic 
environment and to couple the THz radiation via optical ports 
into the cryogenic environment, i.e.~into the fridge. In order to generate 
a large variety of THz radiations, as an example of many other technological directions, 
the quantum cascade laser \cite{Faist1994, Friedli2013} 
may be employed as a very useful device which can additionally be operated 
at liquid nitrogen temperatures. This means that the operation of the quantum 
cascade laser is not limited by technologically 
demanding ultra-low temperature experiments using liquid helium. Also, 
it has been shown, that the quantum cascade laser is suitable for combination 
with waveguide devices in order to controllably guide the generated radiation to free 
space \cite{Justen2017}. Finally, it should be noted that in \cite{Friedli2013}, four-wave mixing 
has been realized in a quantum cascade laser device, i.e.~at least the possibility is suggested
to generate THz radiation with interesting quantum statistics.

\begin{figure}[t]
\includegraphics[width=0.8\columnwidth]{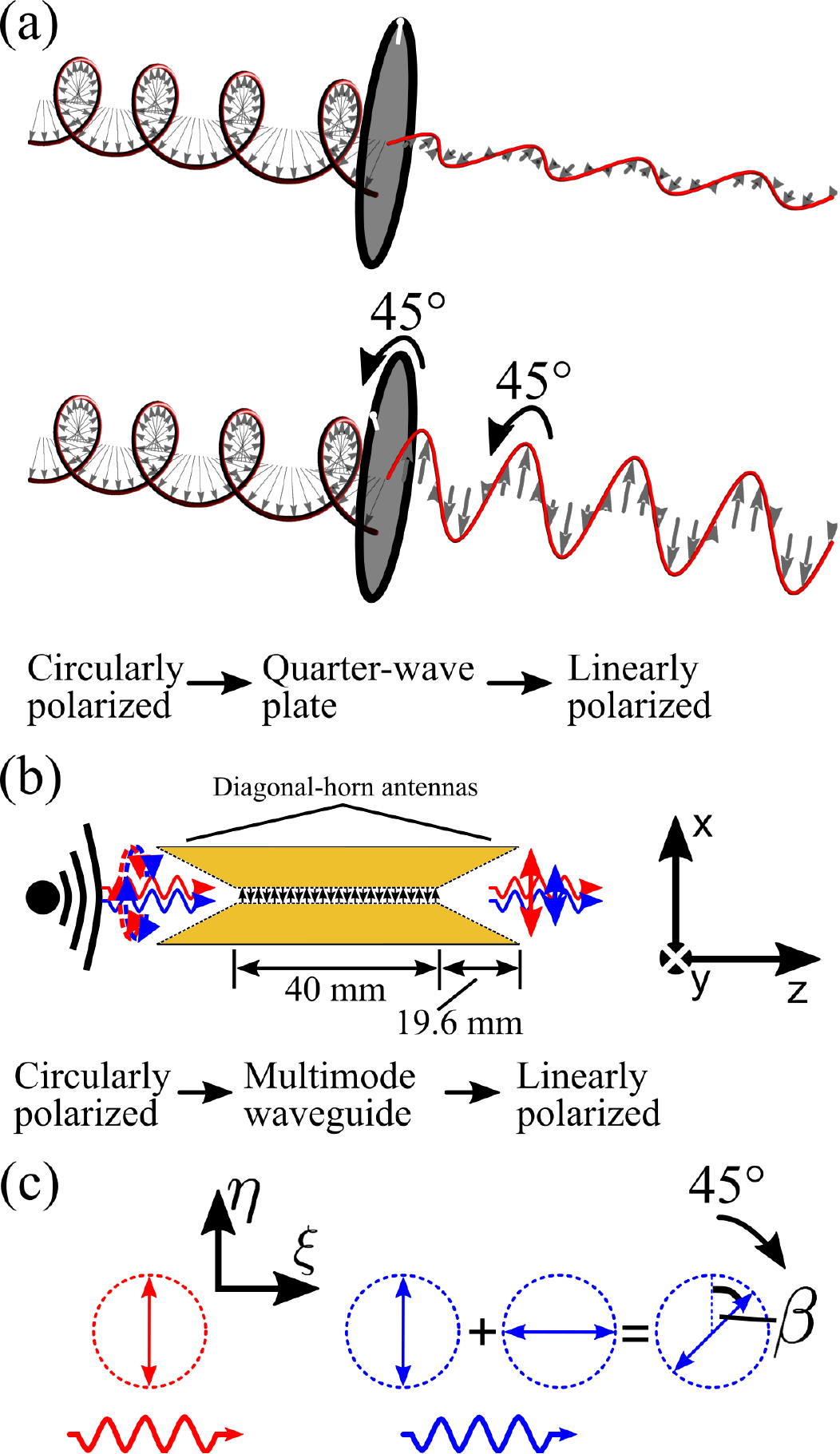}
\caption{\label{fig01}(a) Conventional way of transforming 
a left-circularly polarized electro-magnetic field 
into a linearly polarized field, by rotating {\it mechanically} 
a quarter-wave plate to the desired angle. (b) Multimode 
waveguide and diagonal-horn antenna assembly, acting as a 
{\it non-mechanical} $45^\circ$ polarization rotator, demonstrated 
in this work. The excitation of multiple waveguide modes 
with the circularly-polarized field, distributes the field energy 
equally among the waveguide modes, indicated by the red 
and blue photons. (c) Exciting only the fundamental waveguide 
mode leads to a radiated field from the output diagonal-horn 
antenna with a well defined linear polarization along the 
$\eta$-axis (co-polarization axis) of the aperture coordinate 
system of the diagonal-horn antenna, shown as the red photon. 
Further, exciting higher-order waveguide modes leads to a 
radiated field with a $45^\circ$ rotated polarization due to 
a superposition of linear polarizations along the $\eta$-axis 
and $\xi$-axis (cross-polarization axis), shown as the blue photon. 
}
\end{figure}

With this perspective, the polarization of the THz photon becomes 
a particularly interesting parameter, as it could enable THz photon 
entanglement experiments with the polarization being the 
entanglement parameter. However, in the THz domain it is 
a difficult task to carry out a quantitative measurement of 
the polarization (Fig \ref{fig01}(a)) without significantly 
disturbing the signal. The reason is, that the Rayleigh 
length of optical elements in the THz frequency range 
is usually small, i.e.~only slightly larger than the beam 
waist, such that one often operates at the onset of 
beam divergence.

In the present work we aim at enhancing the toolset 
for THz experiments by establishing a simple method for 
the generation, adjusting and measurement of such a polarization.

We study in experiments and numerical 
simulations the polarization state of a multimode 
free-space sub-THz field as a function of frequency 
in the range from 215 to 580 GHz, that is launched 
from a rectangular waveguide and a diagonal-horn 
antenna, cf.~Fig.~\ref{fig01}(b). We present a method 
to measure the polarization state resulting from these 
multiple modes of the free-space sub-THz field using 
a coherent detector (photo-mixer) in combination with a 
planar silicon-mirror acting as a Fresnel scatterer. This 
enables us to determine the polarization components with 
high accuracy and without the need for any opto-mechanical 
components such as rotatable polarizers. In particular, this 
renders our method suitable for ultra-high frequency cQED 
experiments in a cryogenic environment. 

We find that when only the fundamental TE$_{10}$ 
mode of the waveguide is excited, as expected, the field emitted 
by the diagonal-horn antenna is characterized by a predominantly 
linear polarization. This is consistent with our earlier findings reported 
in \cite{westig2020} whereas the emitted field still contains a 
cross-polarization power component of about 5$\%$. At higher 
frequencies we find in both simulations and experiments 
that excitation of higher-order modes of the waveguide (TE$_{20}$, TE$_{01}$, 
TE$_{11}$ and TM$_{11}$) leads to a well-defined rotation of the 
polarization by up to $45^\circ$, cf.~Fig.~\ref{fig01}(c). 
Despite the higher-order modes, the radiated field appears to maintain 
a predominant Gaussian beam character, since an unidirectional 
coupling to a detector was possible, whereas the unidirectionality 
is independent of frequency.

Section~\ref{sec:02} provides the theoretical basis for the 
experimentally observed polarization rotation effect of the 
emitted THz field, starting from the waveguide theory and, 
eventually, describing the simulation of the near- and far-field 
generation of the THz field by the diagonal horn antenna.
In Section~\ref{sec:03} we describe 
various details of the experiment, including the design of the 
fabricated waveguide and the diagonal horn antennas, as well as 
the functionality of the photo-mixers. Section~\ref{sec:04} 
describes the measurement setup. Section~\ref{sec:05} 
explains the measurement procedure and the method of analysis.
Section~\ref{sec:06} discusses the results. Section~\ref{sec:07} 
concludes our work. Furthermore, we provide a detailed Appendix, 
comprising a description of the calibration of our setup, 
far-field simulations of the output field of the diagonal-horn 
antenna and a discussion of a single-photon detector calibration 
using our waveguide/diagonal-horn antenna device.
\section{\label{sec:02}Polarization rotation 
from waveguide theory and simulation}
\begin{figure*}[tb]
\centering
\includegraphics[width=\textwidth]{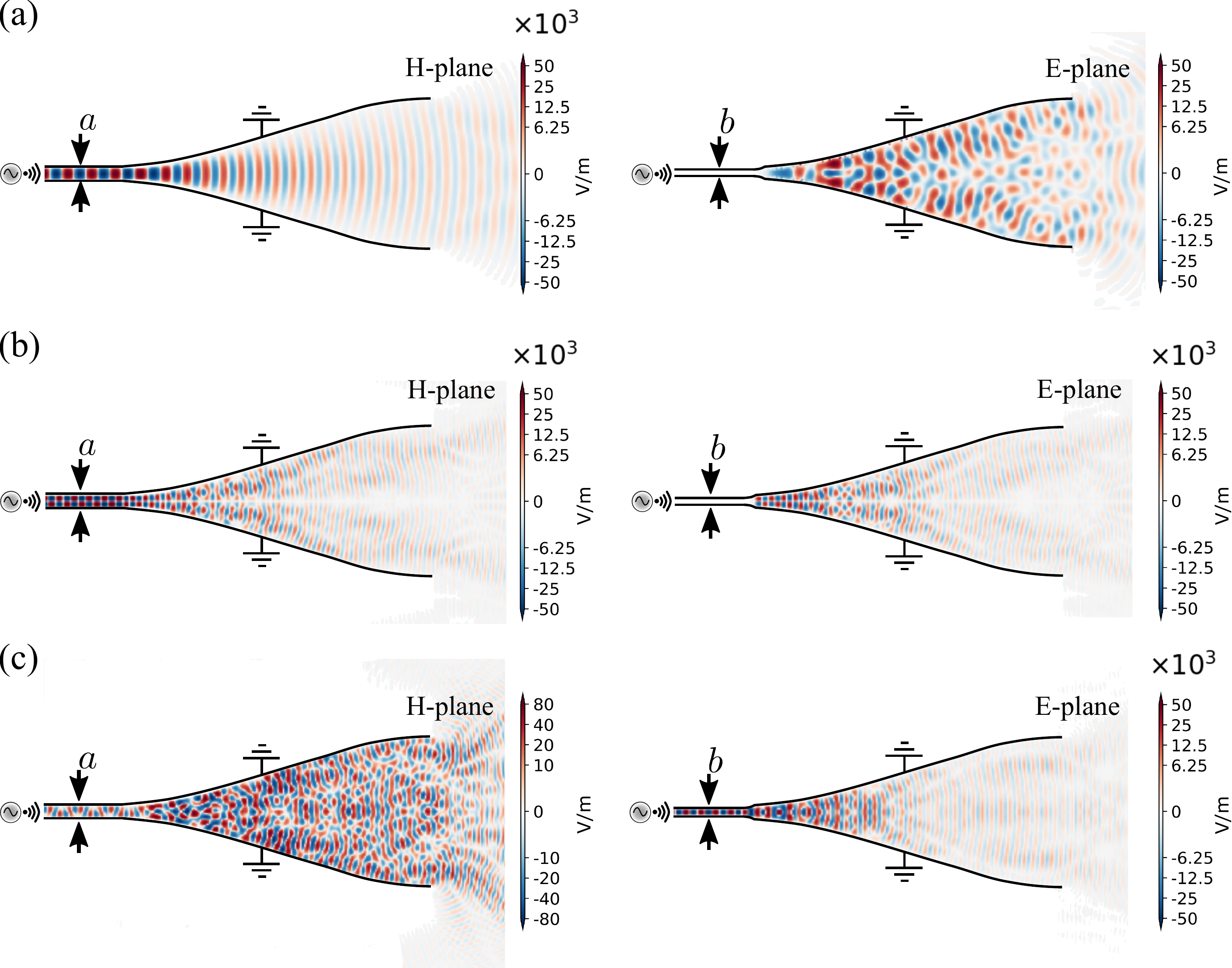}
\caption{\label{fig02}Near-field 
simulation results of the electric portion of the electro-magnetic 
field, expressed as electric field intensity, 
in the experimentally implemented 
antenna-feed waveguide (rectangular section with 
dimensions $a$ and $b$) and bell-mouthed diagonal-horn antenna 
shape, obtained by the CST software \cite{CST}. In the 
antenna-feed waveguide, which is further connected to the 
waveguide channel (cf.~Fig.~\ref{fig03}(b), (c) and (e)), 
a certain mode of the electric portion of the electro-magnetic field 
is excited predominantly in 
the H- or E-plane. After the waveguide excitation, however, 
an electric field develops in both planes of the diagonal-horn 
antenna. The electric field intensities in the diagonal-horn antenna planes 
are mode-dependent. The figure summarizes the results for the 
first three propagating modes at reasonably selected frequencies, 
(a) $\mathrm{TE}_{10}$ (270 GHz), (b) $\mathrm{TE}_{20}$ (480 GHz) 
and (c) $\mathrm{TE}_{01}$ (480 GHz). Note that 
the overall values of the 
electric field intensities, the intensities being 
quantified by the scales in units of V/m, 
in the E- and H-plane of the diagonal-horn antenna for the 
mode $\mathrm{TE}_{10}$, (a), are approximately 
swapped between same E- and H-plane 
for the mode $\mathrm{TE}_{01}$, (c).
The overall values of the electric-field intensities in the 
E- and H-plane of the diagonal-horn antenna are 
approximately the same for the mode $\mathrm{TE}_{20}$, (b). 
With these three modes in superposition, 
polarization rotation by $45^\circ$ occurs around 
the wavevector $k_{z}$, suggested by Fig.~\ref{fig03}(a).
}
\end{figure*}
The electric and magnetic field distributions in all three spatial directions 
($E_{x,y,z}$ and $H_{x,y,z}$) in a rectangular waveguide are fundamentally 
derived as solutions of the reduced wave equations for the electro-magnetic 
field in the waveguide,
\begin{subequations}
\begin{align}
\label{eq:01a}
\left(\frac{\partial^2}{\partial x^2} + \frac{\partial^2}{\partial y^2} + k_{c}^2\right) h_{z}(x,y) = 0
\\
\label{eq:01b}
\left(\frac{\partial^2}{\partial x^2} + \frac{\partial^2}{\partial y^2} + k_{c}^2\right) e_{z}(x,y) = 0~.
\end{align}
\end{subequations}
The $z$-direction is the longitudinal direction of the 
waveguide and the diagonal-horn antenna and at the same time 
the propagation direction of the electro-magnetic field in the 
waveguide and in the diagonal-horn antenna. Furthermore, 
Eq.~(\ref{eq:01a}) obtains solutions for transversal electric fields 
for which $E_{z} = 0$ and Eq.~(\ref{eq:01b}) obtains solutions 
for transversal magnetic fields for which $H_{z} = 0$. Moreover, 
in the above equations, $h_{z}$ fullfils the relation 
$H_{z}(x,y,z) = h_{z}(x,y) \exp{\left(-i\beta z\right)}$ and $e_{z}$ 
fullfils the relation $E_{z}(x,y,z) = e_{z}(x,y) \exp{\left(-i\beta z\right)}$. 
In particular, $k_{c} = \sqrt{k^2 - \beta^2}$ is the cut-off wavenumber 
of the waveguide or a segment of the diagonal horn antenna, 
with $k$ being the wavevector and $\beta$ being the propagation constant.
 
Application of suitable boundary conditions capture the geometry of the 
waveguide and of the diagonal horn antenna.

Specifically, to describe the electro-magnetic field in our 
waveguide/diagonal-horn antenna device, eventually, makes a precise 
description of the transition from the waveguide to the diagonal-horn 
antenna necessary. This transition is a complex mechanical transition, 
cf.~Fig.~\ref{fig03}(e), ii and iii, which renders a precise analytical solution 
of the wave equation Eq.~(\ref{eq:01a}) and (\ref{eq:01b}) unpractical.
 
Additionally, for our purposes, the description of the 
transition from the near- into the far-field for a multimode 
electro-magnetic field, emitted or received by the diagonal-horn 
antenna, is important for the quantitative analysis of our experimental 
results. For a precise evaluation of the far field of the diagonal horn 
antenna, the starting point would be a precisely known aperture field 
which could then be expanded into Gauss-Hermite functions, in order to 
obtain the far-field radiation pattern.

To this end, instead of a fully analytical solution,
we have solved the above equations in a CST \cite{CST} simulation of the 
nominal waveguide and diagonal-horn antenna 
dimensions and shapes, since we expect from this route more 
precise results for our later analysis.

The near-field simulation results are shown in 
Fig.~\ref{fig02} and we specify further in Fig.~\ref{fig03} and 
Sec.~\ref{sec:031} the waveguide and diagonal-horn dimensions. 
Further, we represent E- and H-plane cuts of the electro-magnetic field 
in the waveguide and the diagonal horn-antenna in Fig.~\ref{fig02}. 
In particular, the E- and H-plane are defined on the basis of the fundamental 
mode ($\mathrm{TE}_{10}$) of the waveguide, whereas the E-plane is the 
plane containing the electric-field vector and the H-plane is the plane 
perpendicular to the electric-field vector. The near-field simulation results 
show the projection of the electric field on the E- and H-planes, where red 
colors indicate an electric-field maximum and blue colors indicate an 
electric-field minimum, further characterized by positive and negative 
numbers in units of V/m. Further, since the diagonal horn 
antenna is much larger than the wavelength of the respective 
propagating modes, our simulations evaluate electric fields 
with several electric field maxima (red colors) and electric field 
minima (blue colors), shown in Fig.~\ref{fig02}.

In general, the polarization of 
an electro-magnetic field is defined as the oscillation direction of the 
electric field component of the electro-magnetic field.

Therefore, Fig.~\ref{fig02}(a) shows the expected result for 
the $\mathrm{TE}_{10}$ mode in which the electric field oscillates out 
of the H-plane (red colors) and into the H-plane (blue colors), therefore, 
the polarization is perpendicular to the H-plane and parallel to the E-plane.
Note in particular the scale for the electric field in units of 
$\mathrm{V/m}$, shown for each simulated E- and H-plane cut of the diagonal 
horn antenna shape. In Fig.~\ref{fig02}(a) the predominant electric field is 
established in the H-plane whereas it is approximately three order of 
magnitude smaller in the E-plane. This effect establishes the predominant direction 
of the polarization described before.

In contrast, Fig.~\ref{fig02}(c) suggests that for the $\mathrm{TE}_{01}$ 
mode the polarization is perpendicular to the one of the $\mathrm{TE}_{10}$ 
mode. This is, like described before, evidenced by the electric 
field strength. This time, however, the electric field is established predominantly in the 
E-plane whereas it is approximately three orders of magnitude smaller in the H-plane. Since, 
the E- and H-planes are perpendicular to each other, also the polarization in the 
$\mathrm{TE}_{01}$ is perpendicular to the polarization of the $\mathrm{TE}_{10}$ mode.
Once these two modes propagate simultaneously, the effective polarization is the 
vector sum of the two polarizations of the modes, hence, the polarization is rotated 
by $45^\circ$.

In particular, we show in Appendices~\ref{app:C} and \ref{app:D} that the far-field shows 
corresponding features. This is an important aspect since the far-field is coupled to 
our detection scheme.

The described effects are the fundamental basis of our experimentally observed polarization 
rotation by $\sim 45^\circ$.

Additionally, the higher order modes $\mathrm{TE}_{20}$, 
$\mathrm{TE}_{11}$ and $\mathrm{TM}_{11}$ result in essentially 
near-field patterns of the like shown in Fig.~\ref{fig02}(b). 
Here, the near-field shows a fundamentally different pattern 
compared to the ones in Fig.~\ref{fig02}(a) and (c). At a 
given longitudinal position, on one side of the symmetry axis of 
the diagonal-horn antenna, a maximum or minimum electrical 
field is obtained whereas on the other side of the symmetry axis 
at the same longitudinal position, a respective opposite field is 
found. Additionally, the field intensities in the H- and E-plane are 
practically equal. This means that the effective polarization would 
be zero when averaged over the diagonal-horn aperture. Near-field 
effects of this kind, essentially a capacitive effect due to the 
confined geometry of the diagonal-horn antenna and the waveguide, 
are known to vanish in the far-field. In the far-field, equal amounts of 
power are then obtained in the co- and cross-polarization components 
of the diagonal horn antenna, like we describe in Appendices~\ref{app:C} 
and \ref{app:D} together with far-field simulation results shown in 
Figs.~\ref{figS03}(c) and (d) for the $\mathrm{TE}_{20}$ mode. 
The equal power components in the co- and cross-polarization 
components of the far-field of the diagonal horn antenna for the 
modes $\mathrm{TE}_{20}$, $\mathrm{TE}_{11}$ and 
$\mathrm{TM}_{11}$, lead to individual propagating fields with 
$\sim 45^\circ$ rotated polarization. When all five modes 
propagate, a superposed electro-magnetic field is obtained, 
being characterized by the aforementioned rotated polarization.

Importantly, for the far-field we studied in our 
simulations the higher-order propagating modes in the 
waveguide and the created multimode (in our case 
up to five) electro-magnetic field. Interestingly, we have
discovered that this field is characterized by 
co- and cross polarization components of the electric field 
which are practically in-phase in the same mode, 
when radiated from the diagonal-horn antenna into 
free space. Furthermore, our multimode simulations 
find that the phase-delay between the aforementioned far-field 
(i.e.~free space) electric fields in different modes is 
practically negligible as well. This means, that 
the different modes are emitted by the diagonal-horn 
antenna in a coherent fashion and are practically not 
time-delayed with respect to each other. This 
is key for an effective polarization rotation of 
$45^\circ$ to happen and for generating a coherent 
electric field with a predominant linear polarization 
content and with a negligible circular polarization content.
\section{\label{sec:03}Experimental system}
\subsection{\label{sec:031}Waveguide assembly}
\begin{figure*}[htbp]
\centering
\includegraphics[width=\textwidth]{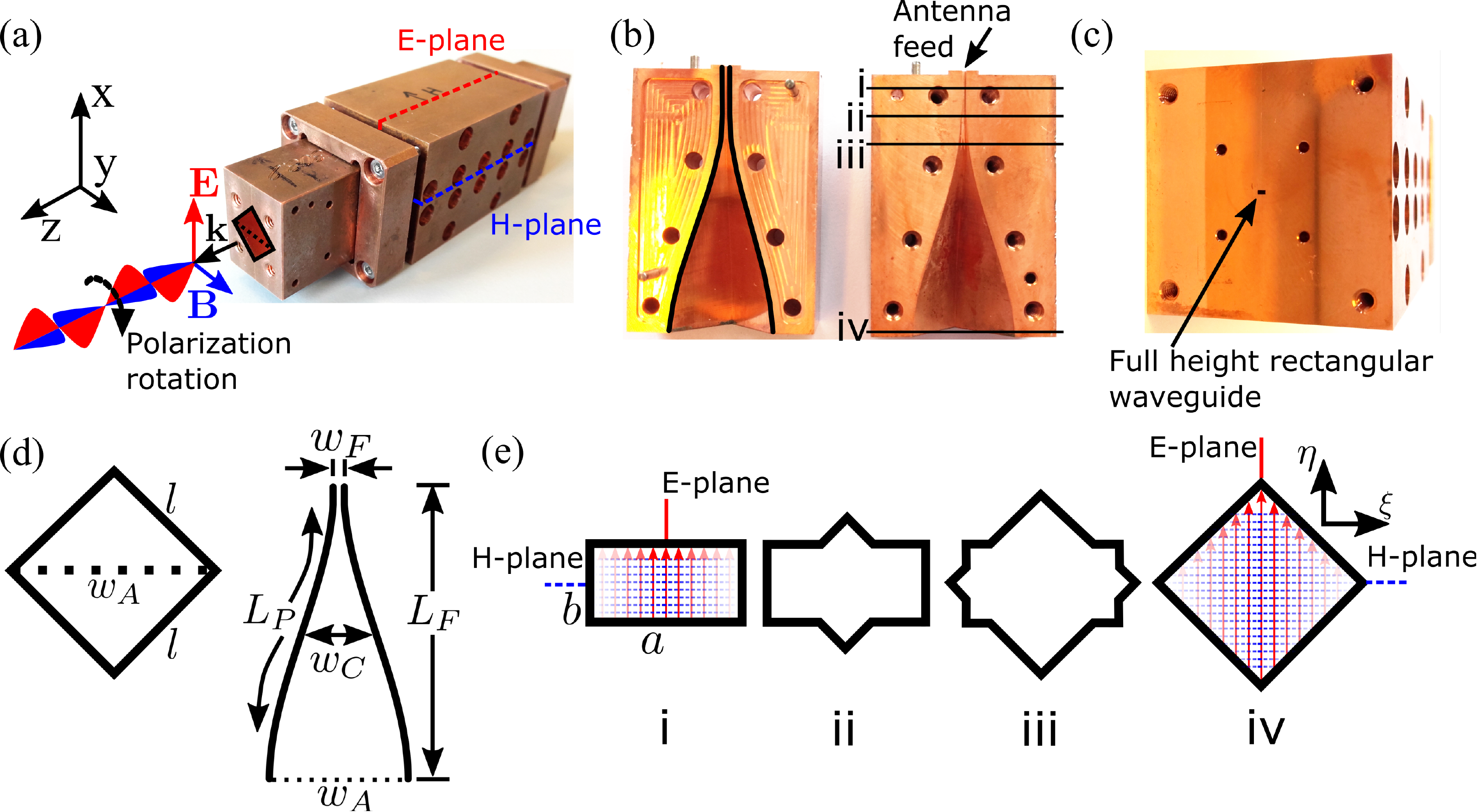}
\caption{\label{fig03}Design, components and assembly of 
the fabricated diagonal-horn antenna and waveguide device. (a) The completed 
assembly, diagonal-horn antennas opened along the E-plane (b) which are connected by 
a full-height rectangular waveguide (c). 
In (b), left: the black profile emphasizes the antenna profile. Right: 
subdivision into cross sections (i)-(iv), shown in more detail in (e).
(d) Cross-sections of the diagonal-horn 
antenna aperture (left) and the profile (right). (e) Cross-sections indicated in 
(b), showing the inner conductor-shape of the 
waveguide and the diagonal-horn 
antenna, with the electro-magnetic field of the fundamental 
$\mathrm{TE}_{10}$-mode in the waveguide 
and diagonal-horn antenna at positions i and iv. 
The $\eta$-$\xi$ aperture coordinate system is shown in iv. 
}
\end{figure*}

In order to test the aforementioned prediction, our starting point 
is a machined diagonal-horn antenna and waveguide 
assembly as shown in Fig.~\ref{fig03}(a), suitable for the 
frequency range from 215 to 580~GHz. Similar units are commonly 
used in mixer-assemblies for heterodyne detection in astronomical 
instruments (see for example \citet{HIFI}). It is made from the material 
CuTe, with waveguide dimensions (Fig.~\ref{fig03}(c) and (e)) of  
$a = 800~\mu\mathrm{m}$ and $b = 400~\mu\mathrm{m}$. At each 
end a diagonal-horn antenna is attached with the feedpoint of the 
two antennas matching the waveguide dimensions. 

Figure~\ref{fig03}(a) shows a completed unit and a 
representation of an emitted electro-magnetic field. The 
diagonal horn can be disassembled into two halves along its 
E-plane (Fig.~\ref{fig03}(b)), which reveals its dimensions as 
defined in Fig.~\ref{fig03}(d). The dimensions in Fig.~\ref{fig03}(d) 
of the diagonal-horn antenna aperture (left) and the profile (right) 
are $w_{A} = 9.9~\mathrm{mm}$ (geometric aperture-width), 
$l = 7~\mathrm{mm}$ (geometric aperture edge-length), 
$w_F = 400~\mu\mathrm{m}$ (waveguide feed, $b$-side,  
$L_F = 19.6~\mathrm{mm}$ (feed length), 
$L_P = 21.48~\mathrm{mm}$ (profile length) and 
$w_C = 4.91~\mathrm{mm}$ (width of the horn profile at 
distance $L_{F}/2$ from the feedpoint). 

The cross-sections shown in Fig.~\ref{fig03}(e) 
picture the electro-magnetic field of the fundamental 
$\mathrm{TE}_{10}$-mode in the waveguide (Fig.~\ref{fig03}(b) 
and (c)) at position  i, and in the diagonal-horn antenna aperture 
at position iv. In the last panel of Fig.~\ref{fig03}(e), 
we depict the aperture coordinates $\eta$ and $\xi$, 
already introduced in Fig.~\ref{fig01}(c). The co- and 
cross-polarizations point in direction of $\eta$ and $\xi$.
  
The signal path in our setup is as follows. The input diagonal-horn antenna 
receives an electro-magnetic sub-THz field generated by a photo-mixer, exposed to the 
signal of two coupled DFB lasers. This signal excites the waveguide with the multimode 
field. Subsequently, the waveguide excites the output diagonal-horn antenna which then 
emits a multimode electro-magnetic field into free space where it 
gets reflected from the planar silicon-mirror towards a coherent 
detector, where it is detected and analyzed.
\subsection{\label{sec:032}Photo-mixers}
To build the foundation for a better understanding of our 
experiments, we will describe in more detail the way of operation of the 
photo-mixers by means of our experimental setup, shown 
in Fig.~\ref{fig04}. The frequency-tunable electro-magnetic signal, in the 
frequency range of 215 to 580~GHz, is generated and detected by 
superimposing the outputs of two 780~nm distributed feedback (DFB) 
lasers in a beam combiner (BC) and shone on two GaAs photo-mixers 
connected at the output of the beam combiner via polarization 
maintaining fibers (PMF). One photo-mixer acts as coherent 
sub-THz source (S) and the second acts as a coherent 
sub-THz detector (D) \cite{toptica}. The incident laser power 
on each photo-mixer is approximately $30~\mathrm{mW}$. The desired 
frequency of the sub-THz electro-magnetic field is set by adjusting the 
difference frequency, $f$,  between the two DFB lasers. Optimal coupling 
between all optical elements is achieved by arranging the set-up 
in such a way that the propagating Gaussian beam-divergence is minimized 
and a narrow beam hits the detector.

The planar-silicon mirror at the output makes it possible to measure 
the polarization by using Fresnel scattering, to be discussed below, without 
using any movable parts and with a minimal number of optical components.  
\section{\label{sec:04}Measurement Setup}
\begin{figure*}[htbp]
\centering
\includegraphics[width=\textwidth]{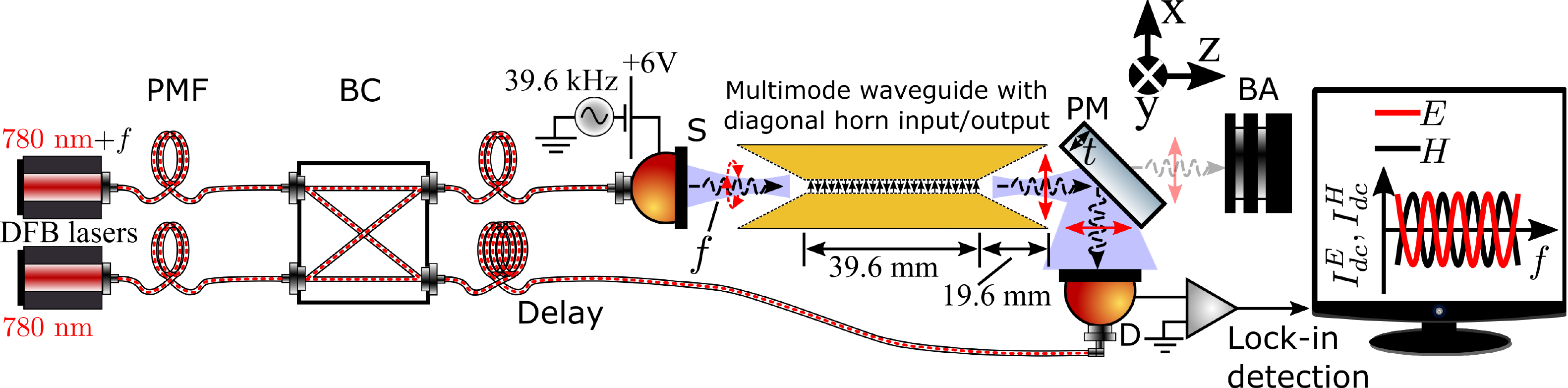}
\caption{\label{fig04}
Measurement system to measure the polarization at the output 
of the diagonal-horn antenna. Coherent sub-THz source (S) and 
detector (D) are used and the lock-in detected signal is displayed as a function 
of frequency. The planar silicon-mirror (PM, TYDEX \cite{tydex}) 
acts as a Fresnel scatterer and enables the 
polarization measurement. It reflects linearly-polarized 
electric fields with polarization components parallel and 
perpendicular to the plane of incidence (paper plane), with opposite 
signs (and slightly different magnitudes) into the aperture 
of the coherent detector. Two successive 
measurements result in the detector currents 
$I_{dc}^{E}$ and $I_{dc}^{H}$, Eqs.~(\ref{eq:03a}) and (\ref{eq:03b}), 
when the E- and H-plane of the waveguide (cf.~Fig.~\ref{fig03}) 
and diagonal-horn antenna are successively aligned parallel to the 
plane of incidence. This implements a direct measurement of 
the correlation function in Eq.~(\ref{eq:04}). A beam absorber (BA) 
attenuates a standing wave in the setup.}
\end{figure*}

The polarization rotation is measured based on the principle of Fresnel scattering, 
implemented by the scattering of the electro-magnetic field
from a silicon mirror, cf.~Fig.~\ref{fig04}. The scattered field is received by the 
detector with different signs, because either the positive or negative region of 
the electro-magnetic field oscillation reaches the detector 
area first. Equivalently, this corresponds to a phase shift 
of $\pi$ of the electro-magnetic field which 
depends on the linear polarization of the field, according to 
the Fresnel theory \cite{born1985}. The sign is measured directly in 
our coherent detection scheme, since it determines the 
dc-photocurrent direction. 

The signal is detected by coherent detection of the scattered 
electro-magnetic field, which contains the phase information 
for the two different waveguide orientations (cf.~caption of 
Fig.~\ref{fig04}), which we adjust in two successive measurements. 
The phase information of the detected 
field is extracted by post-processing the frequency-dependent 
transmission data between source and detector in 
Fig.~\ref{fig04} by means of a Hilbert transformation 
\cite{roggenbuck2010, westig2020}, described below. In this way we build the basis 
for the evaluation of the phase-phase correlation function [Eq.~(\ref{eq:04})] for the 
two different waveguide orientations.

Two details are important in the interpretation of our 
measurement results. First, the direction of 
the current flow in the detector. The dc-photocurrent $I_{dc}$ is 
periodic with the detuning frequency $f$ and dependents on the delay length 
$\Delta L = L_{S} + L_{0} - L_{D}$ between 
the optical fibers, including the free space path of 
the sub-THz field from the source to the 
detector \cite{roggenbuck2010}. 
$L_{S}$ and $L_{D}$ are the (different) optical path 
lengths travelled by the two superimposed DFB laser fields to the source and 
detector through the optical fibers. The length $L_{0}$ is the 
additional path length, travelled by the sub-THz 
field from the source to the detector through free space, 
through the diagonal horns and through the waveguide (black 
wiggly and dashed arrows in Fig.~\ref{fig04}). 

Secondly, the sign and the magnitude of $I_{dc}$ 
is also determined by two more sets of parameters related to the 
Fresnel scattering effect. The first parameter set is the sign and the absolute 
value of the Fresnel amplitude reflection coefficient, $r^{\perp, \parallel}(f)$, 
of the electro-magnetic field at the output of the planar silicon-mirror, where 
the electric field component has a polarization perpendicular 
to ($\perp$) or/and parallel ($\parallel$) to the 
planar silicon-mirror plane of incidence. Furthermore, the sub-THz 
electro-magnetic field has a well defined phase $\varphi^{\perp}$ or 
$\varphi^{\parallel}$ for each of the two polarizations. 
Generally, these phases dot not have the same values, 
but are in practice not much shifted with respect to each other. The second 
parameter is the amplitude of the electric component, contained in the two 
polarization components perpendicular ($\perp$) 
and/or parallel ($\parallel$) with respect to the planar silicon-mirror 
plane of incidence. In this work it is sufficient to determine the phases 
of the output field for two different orientations of the waveguide. 

The polarization angle of 
the output field of the diagonal-horn antenna follows now from a 
statistical analysis by means of a phase-phase 
correlation function [Eq.~(\ref{eq:04})]. The idea is that the phase 
of the output field after scattering from the planar silicon-mirror into 
the detector differs by a shift of exactly $\pi$ between the two 
successive measurements, which signals the detection of pure 
$\perp$- and $\parallel$-components. This $\pi$-shift of the phase 
is well known and described by the Fresnel theory \cite{born1985}, 
but it should be supplemented with another phase shift due to the 
finite thickness of the planar silicon-mirror. 
For the real-valued detector currents $I_{dc}(f)$, flowing in 
response to a detected electro-magnetic field of frequency $f$, 
the analytical complex-valued detector current reads
\begin{equation}
\label{eq:02}
\mathcal{I}_{dc}(f) = I_{dc}(f) + i\mathcal{H}\left[I_{dc}(f)\right] 
= S(f)\exp\left[i\phi(f)\right]~.
\end{equation}
Here, $\mathcal{H}(\cdots)$ is the Hilbert transformation \cite{vogt2017}, 
$\phi(f)$ the instantaneous phase of the signal and $S(f)$ 
is the instantaneous amplitude. For the rest of the paper, $\phi(f)$ is the 
key observable from which we derive our results, explained in more 
detail below. 

A phase shift deviating from $\pi$, eventually, resulting in 
phase jumps in the detector, is expected for an output field characterized 
by a mixture of $\perp$- and $\parallel$-components. We explain this case 
in more detail in the following paragraph, in particular the case of an equal 
mixture of $\perp$- and $\parallel$-components, hence, an output field 
with $45^\circ$ rotated polarization.
\section{\label{sec:05}Measurement procedure and method of analysis}
\subsection{\label{sec:051}Obtaining the data}
The sub-THz electric field component received by the 
detector leads to an ac-voltage drop across an interdigitated 
capacitor part of the detector with a frequency equal to the difference in laser frequencies 
$f$. Together with the laser-induced impedance 
modulation at the same frequency, but in general with a 
different phase, a coherent dc-photocurrent, $I_{dc}(f)$, 
flows in the positive or negative direction (dependent on 
the phase) across the feedpoint of the log-spiral circuit.
We detect this dc-photocurrent with a post-amplification scheme 
described in \cite{westig2020},  with each data point 
integrated over 500~ms. This detection scheme 
resembles a coherent detector at sub-THz frequencies with a 
high-dynamic range up to 80~dB \cite{toptica} like described 
by \citet{roggenbuck2010}. A beneficial aspect of this scheme is that 
it measures the transmitted amplitude rather than 
only the transmitted power. This allows us to use the 
planar silicon-mirror in our setup as a Fresnel scatterer. 

We perform our measurements in two 
successive steps. First, we align the waveguide and 
diagonal-horn antenna with the E-plane parallel to the plane 
of incidence and, secondly, we align 
them with their H-plane parallel to the plane of incidence. 
For each of these steps we record the detector current 
(given in analytical form 
in Eqs.~(\ref{eq:03a}) and (\ref{eq:03b}))
as a function of frequency, covering the range of 215~GHz to 580~GHz.
For the fundamental waveguide mode up to a frequency of 
about 400~GHz, determined by the diagonal-horn antenna and 
the rectangular waveguide geometry \cite{westig2020}, the polarization is 
predominantly parallel to the E-plane. By rotating the rectangular 
waveguide and diagonal-horn antenna by 90 degrees, 
we also rotate the polarization by the same amount. By adding 
the planar silicon-mirror to the setup described in \cite{westig2020}, 
we obtain the polarization sensitive coherent detector.

In the measurement situation in which the rectangular
waveguide and diagonal-horn antenna assembly is aligned such that the 
E- or H-plane is parallel to the silicon-mirror plane of incidence,
we can express the detector currents as,
\begin{widetext}
\begin{subequations}
\begin{align}
\label{eq:03a}
I_{dc}^{E}(f) &= \mathcal{C} \sum_{l} 
\mathcal{A}\cdot\mathrm{Re}\left[r^{\parallel}(f)\right] 
\mathcal{E}_{l}^{E}\cos\left[\frac{2\pi f \Delta L}{c_{0}} + 
\varphi^{\parallel}_{l} + \Delta\varphi^{\parallel - \perp}_{l}\right]C^{(1)}_{l}(f) + \mathrm{Re}\left[r^{\perp}(f)\right] 
\mathcal{E}_{l}^{H}\cos\left[\frac{2\pi f \Delta L}{c_{0}} + 
\varphi^{\perp}_{l}\right]C^{(2)}_{l}(f)~,
\\
\label{eq:03b}
I_{dc}^{H}(f) &= \mathcal{C} \sum_{l} \mathrm{Re}\left[r^{\perp}(f)\right] 
\mathcal{E}_{l}^{E}\cos\left[\frac{2\pi f \Delta L}{c_{0}} + 
\varphi^{\parallel}_{l}\right]C^{(3)}_{l}(f) + \mathcal{A}\cdot\mathrm{Re}\left[r^{\parallel}(f)\right]
\mathcal{E}_{l}^{H}\cos\left[\frac{2\pi f \Delta L}{c_{0}} + 
\varphi^{\perp}_{l} + \Delta\varphi^{\parallel - \perp}_{l}\right]C^{(4)}_{l}(f)~.
\end{align}
\end{subequations}
\end{widetext}
The amplitudes $\mathcal{E}^{E,H}_{l}$ quantify the 
field strength in the E- or H-plane for a given mode 
$l \in \lbrace 0\ldots5 \rbrace$ (for more details 
cf.~Appendix~\ref{app:C} and \ref{app:D}). Up to a constant they fully 
determine the size of the detector current. When multiplied 
with the Fresnel scattering amplitudes $r^{\parallel, \perp}$ 
and up to a propagation factor and the polarization 
orientation in free space, the resulting expression is 
equivalent to free space propagating fields 
$\boldsymbol{\mathcal{E}}_{\parallel}$ or 
$\boldsymbol{\mathcal{E}}_{\perp}$. Furthermore,  
$c_{0}$ is the velocity of light in vacuum and 
$\mathcal{C} = d Z_{0}/2Z_{det}^{2} 
\approx 1\cdot10^{-12} \mathrm{-} 1\cdot10^{-13}~\mathrm{m/\Omega}$ 
is the coupling constant between the free space electro-magnetic field 
and the detector which we assume to be the same for the 
$\parallel$- and $\perp$-components. 
Once the detector is fixed on 
the optics table and its position cannot be optimized anymore for 
maximal response, the coupling to the detector will be different for the 
$xz$- and the $xy$-plane due to imperfections of the detector. 
We account for this asymmetry by the constant $\mathcal{A} = 0.85$, 
which we determine experimentally by measuring the Fresnel amplitude 
reflection coefficients for the $\perp$- and the $\parallel$-components 
(for more details cf.~Appendix~\ref{app:B}).

The real part of each Fresnel amplitude reflection 
coefficient in Eqs.~(\ref{eq:03a}) and (\ref{eq:03b}) 
contributes in two ways to the measured dc-photocurrent. 
First, it evaluates the sign (or equivalently the phase shift) 
of the scattered wave and, second, it quantifies the frequency 
dependent reflection of the electro-magnetic field from the 
planar silicon-mirror.

The argument of the cosine, $2\pi f \Delta L / c_{0}$, describes the 
frequency periodicity of the detected field when it arrives with a certain 
time delay $\Delta L /c_{0}$ at the detector, as described before.
The phase shifts of the detected polarization components, 
$\phi^{\parallel, \perp}$, add in a similar fashion to the argument 
of the cosine. In our modeling we 
equate the phase difference between the in Sec.~\ref{sec:061} 
further discussed co- and cross-polarizations with $\phi^{\parallel}$ 
while setting $\phi^{\perp} = 0$. Further details are shown 
in Fig.~\ref{figS04}(b) and (d) of Appendix~\ref{app:D}. 

We account also for the imaginary part of the Fresnel 
amplitude reflection coefficients upon scattering from the planar 
silicon-mirror by evaluating their difference in phase shift between 
$\parallel$- and $\perp$-components, $\Delta \phi_{l}^{\parallel - \perp}$.
The fundamental reason for this extra phase shift is the finite thickness 
of the planar silicon-mirror which imposes different phase shifts on the 
$\parallel$- and $\perp$-components when scattered to its output.

Finally, the current correction coefficients $C^{i}_{l=0}(f) = 1$ (single mode) and for $l>1$ 
(multiple modes), $C^{i}_{l}(f) \in [0,1]$ are unknown. Nevertheless, these 
coefficients shall provide the basis for corrections to the detector current. 
Such a correction seems necessary, 
because for the case $l>1$, multiple currents flow in parallel in the active 
detector area while one measures only the resulting (sum) effective current. 
In the standard photo-mixer theory, 
the theoretical framework of multiple-mode detection is not largely discussed 
and no solution seems to be prepared so far. 

\begin{figure*}[htbp]
\centering
\includegraphics[width=\textwidth]{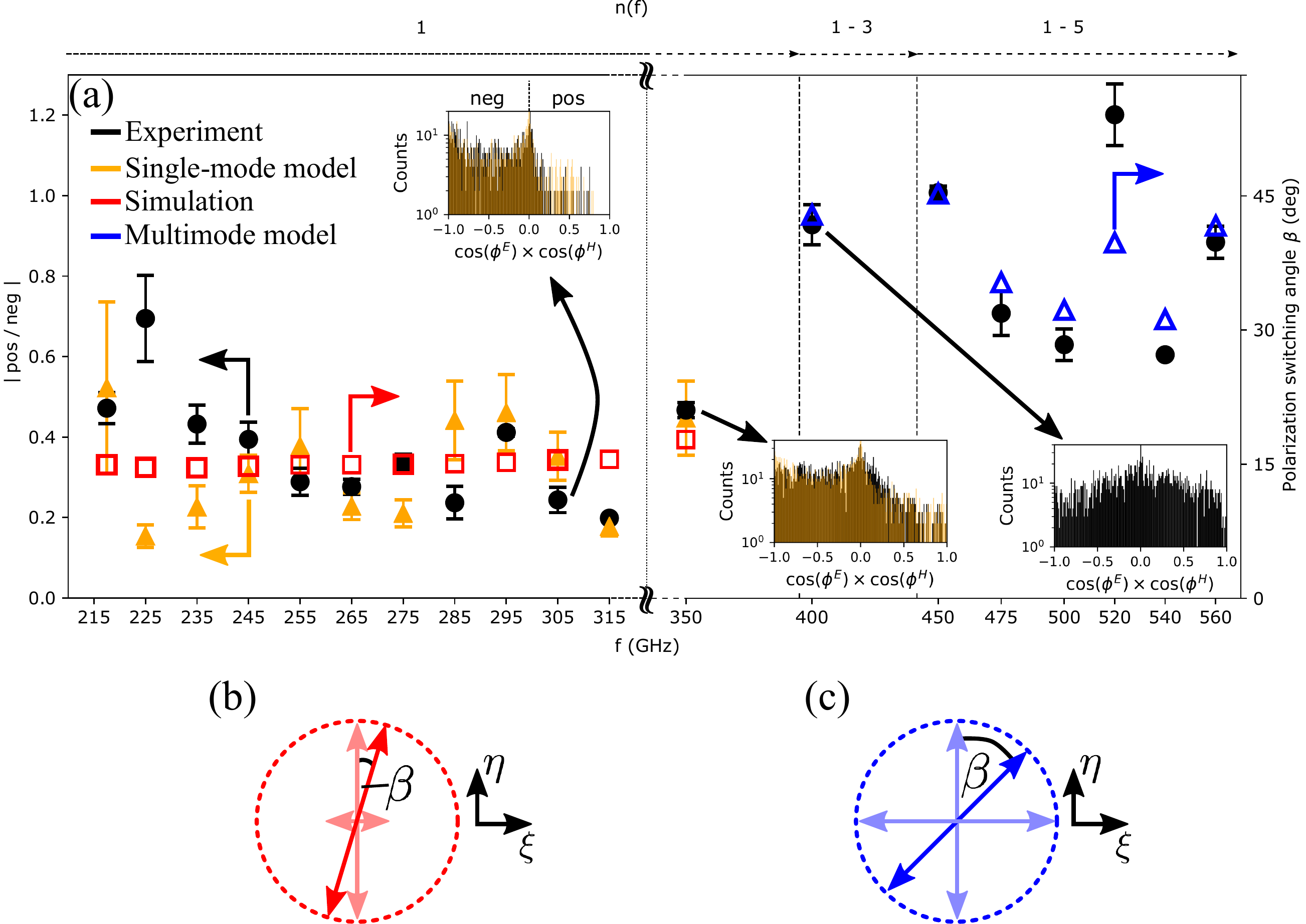}
\caption{\label{fig05}Overview of the results on the measured polarization 
of the rectangular waveguide and diagonal-horn antenna assembly over 
the frequency range from 
215 to 580 GHz. The line at the top indicates the presence of the number 
of higher order modes. Note the scale change at 315 GHz. Panel (a) 
Experimental (black) and simulation (red) results together 
with results for the single-mode (orange) and multimode (blue) 
model. The experimental data-points (black) are evaluated from 
histograms like shown in the inset, by dividing the positive 
counts by the negative ones, \emph{pos/neg} (left y-axis). The 
black-colored histograms contain the experimental values of the 
instantaneous phase-phase correlation function 
for the output electro-magnetic field, $\cos(\phi^{E}) \times \cos(\phi^{H})$ 
(Fig.~\ref{fig04} and Eq.~(\ref{eq:04})).}
\end{figure*} 

\subsection{\label{sec:052}Data analysis}
In order to extract the polarization content from the 
measured detector responses, contained in the phases 
of Eqs.~(\ref{eq:03a}) and (\ref{eq:03b}), we need to perform 
a statistical analysis on these instantaneous phases by means 
of correlation functions.  The Hilbert transformation, Eq.~(\ref{eq:02}), 
evaluates the instantaneous 
phases $\phi^{E}(f)$ or $\phi^{H}(f)$. Each of these phases as a function 
of frequency can be selected by orienting the rectangular
waveguide and diagonal-horn antenna assembly with its E- or 
H-plane parallel with respect 
to the silicon-mirror plane of incidence. 

In the experimental data, the origin of instantaneous phase values 
and most dominant contributions 
are hidden. However, from Eqs.~(\ref{eq:03a}) 
and (\ref{eq:03b}) a number of different contributions to the phase shift are obvious. The most dominant terms are of the form 
$\mathrm{Re}\left[r^{\parallel, \perp}(f)\right] \mathcal{E}_{l}^{E}$ 
and $\mathrm{Re}\left[r^{\parallel, \perp}(f)\right] \mathcal{E}_{l}^{H}$ 
and are the ones which are due to the Fresnel scattering.  
Finally, a correlator of the form 
\begin{equation}
\label{eq:04}
C\left(\phi^{E},\phi^{H}\right) = \cos\left(\phi^{E}\right) 
\times \cos\left(\phi^{H}\right)~,
\end{equation}
yields the phase-phase correlation function of the instantaneous 
phases. In particular, 
Eq.~(\ref{eq:04}) evaluates to '1' when the instantaneous 
phases of the E- and H-plane are in-phase 
and it evaluates to '-1' when they are out-of-phase, i.e.~shifted by $\pi$ with 
respect to each other. Such a shift is expected for 
an ideal linear polarization due to the Fresnel scattering.  Continuous values between '1' and '-1' are possible as well 
and quantify some extra phase shifts which can occur. These extra 
phase shifts have as a source the terms $\varphi_{l}^{\parallel,\perp}$ 
and $\Delta \varphi^{\parallel-\perp}$ in Eqs.~(\ref{eq:03a}) 
and (\ref{eq:03b}). They are usually small, i.e.~influencing the 
measurement results only in a range smaller than the error bars in Fig.~\ref{fig05}, compared to a
more dominant effect, occurring when two similarly large orthogonal polarizations scatter off the planar 
silicon-mirror and are detected at the same time. This drives positive as well as negative 
detector currents (cf.~Appendix~\ref{app:A}) which tend to cancel each other, leading to 
phase jumps and continuous 
correlator values between '1' and '-1'. This is also the expected experimental signature 
of the $45^{\circ}$ linear-polarization rotation, cf.~Fig.~\ref{fig01}(c). 
For the case of a linear polarization, containing just a small cross-polarization 
component, one expects a different distribution of correlator values '1' and '-1' 
compared to a $45^{\circ}$ linear-polarization rotation. In the former 
case, mostly values of '-1' should be obtained because of the 
smallness of the cross-polarization component. We confirm this outcome 
consistently in our experiment.

It is beneficial to understand the experimental 
data, by evaluating the correlator in Eq.~(\ref{eq:04}) over a meaningful 
frequency bandwidth in the measured frequency range. For measurements 
exciting only the fundamental $\mathrm{TE}_{10}$ mode, we evaluate the 
correlator over a bandwidth of 10~GHz to obtain a single data-point 
and for measurements which excite higher-order modes, we have 
chosen a bandwidth of 20~GHz. Through this choice a large 
enough sample of correlator values can be used 
to compare with the theoretical model. In addition, it has proven to 
be convenient to quantify the correlator by plotting its 
values in a histogram, as shown in 
Fig.~\ref{fig05}. 

\section{\label{sec:06}Discussion of the results}
In our experiment, we have measured the multimode field from the output 
diagonal-horn antenna through the detector currents 
$I_{dc}^{E}(f)$ and $I_{dc}^{H}(f)$ as a function of frequency. 
The detector currents are modeled by the 
analytical form of the Eqs.~(\ref{eq:03a}) and (\ref{eq:03b}). 
The statistical analysis of the detector 
currents leads to histograms like shown in the inset of Fig.~\ref{fig05}(a). They 
contain the value distribution of the phase-phase correlation function, 
Eq.~(\ref{eq:04}). In a next step we sum over the positive and negative 
counts in the histograms and build the quotient {\it pos/neg}. 
This is shown as the black data-points which refer to the left part of the 
y-axis in Fig.~\ref{fig05}(a).

In order to relate this measurement to the polarization angle $\beta$ 
(shown in Fig.~\ref{fig05} and also in Fig.~\ref{fig01}(c)), 
we combine our measurements with 
electro-magnetic field simulations of the radiation pattern from 
the exact diagonal-horn antenna geometry. From these simulations, 
we obtain the electric-field strengths $\boldsymbol{\mathcal{E}}_{\eta}$ 
and $\boldsymbol{\mathcal{E}}_{\xi}$, which the diagonal-horn antenna 
radiates into the far-field with the rectangular waveguide acting as the excitation source. 
The polarization angle is given by
\begin{equation}
\label{eq:05}
\beta = \arctan{\left(\boldsymbol{\mathcal{E}}_{\xi}
/\boldsymbol{\mathcal{E}}_{\eta}\right)}~.
\end{equation} 
We use the computer-aided 3D mechanical design 
of the diagonal-horn antenna to model the exact antenna geometry in the 
electro-magnetic field simulation software CST \cite{CST}. 
\subsection{\label{sec:061}Fundamental $\mathrm{TE}_{10}$ mode}
First, we discuss the results for the 
fundamental $\mathrm{TE}_{10}$ mode and address later the multimode case. 
When exciting in the simulation this mode, propagating over the range from 
180~GHz to 360~GHz, we obtain at each selected frequency a set of 
electric-field strengths $(\boldsymbol{\mathcal{E}}_{\eta}, 
\boldsymbol{\mathcal{E}}_{\xi})$. They quantify the far-field 
radiation pattern and through this also the direction of the 
polarization, illustrated in Fig.~\ref{fig05}(b). 
This is shown as the red data points in Fig.~\ref{fig05}(a), which 
are consistent with a 5\% cross-polarization power 
component of the diagonal-horn antenna. For more details on 
the frequency dependence of the cross-polarization we refer 
to Fig.~\ref{figS04}(a) of Appendix~\ref{app:D}. 
Note that the field strengths $(\boldsymbol{\mathcal{E}}_{\eta}, 
\boldsymbol{\mathcal{E}}_{\xi})$ refer to the aperture coordinate 
system of the diagonal-horn antenna, i.e.~they are fixed
to the frame of reference of the diagonal-horn antenna and 
have to be distinguished from the components 
$(\boldsymbol{\mathcal{E}}_{\parallel}, \boldsymbol{\mathcal{E}}_{\perp})$, 
which refer to the planar silicon-mirror plane of incidence. 
More details are given in Appendix~\ref{app:D}. Since the simulation 
results $(\boldsymbol{\mathcal{E}}_{\eta}, \boldsymbol{\mathcal{E}}_{\xi})$ 
are complex valued, we also obtain the phase information of the orthogonal 
field components. For more details on the frequency dependence of 
this phase we refer to Fig.~\ref{figS04}(b) of Appendix~\ref{app:D}. Together with the electric-field strength 
we, therefore, fix for the detector currents every free parameter in Eqs.~(\ref{eq:03a}) and 
(\ref{eq:03b}). Finally, by substituting the values 
obtained from the simulation in the $\eta \textrm{-}\xi$ aperture 
coordinate system into  Eqs.~(\ref{eq:03a}) and (\ref{eq:03b}), we 
need to determine which field component lies in the E- or H- plane. 
For the $\mathrm{TE}_{10}$ mode, the principal field direction is along 
the E-plane. Consequently, 
$\lvert \boldsymbol{\mathcal{E}}_{\eta} \rvert \widehat{=} \mathcal{E}^{E}$, 
$\lvert \boldsymbol{\mathcal{E}}_{\xi} \rvert \widehat{=} \mathcal{E}^{H}$ 
and corresponding substitutions hold for the phases.
A Hilbert transformation of the obtained 
Eqs.~(\ref{eq:03a}) and (\ref{eq:03b}) provides the instantaneous 
phases $\phi^{E}$ and $\phi^{H}$ and the correlator $C(\phi^{E},\phi^{H})$, 
Eq.~(\ref{eq:04}). By this procedure we obtain the orange-colored histograms 
and data points in Fig.~\ref{fig05}(a). In order to obtain the latter, 
we sum again over the positive and negative counts in the model histograms. 

An exact match with the experimental data is not obtained, which 
is not surprising given the complexity of the experiment.
However, the key features are correctly described by our model. 
For the frequencies 285~GHz to 350~GHz, the trend of the data is 
correctly predicted and the absolute values of the experiment and the 
single-mode model are close to each other. In the frequency range 
245~GHz to 275~GHz, a comparable trend of the model and the experiment 
is not evident, but the absolute values are again close to each other. 
Furthermore, the obtained orange-colored model histograms compare 
sufficiently well to the black-colored experimentally determined histograms. 

In particular, we like to highlight
the matching shapes between the experimentally determined histograms 
and the model histograms at 305~GHz and 350~GHz. The histogram at 
305~GHz shows predominantly negative values of the correlator 
$C(\phi^{E},\phi^{H})$. This is indicative of a predominantly linearly 
polarized electro-magnetic field, as explained in Sec.~\ref{sec:04} and 
Sec.~\ref{sec:052}. Moreover, a distribution of negative values and a 
few positive values is obtained for the correlator. This signifies that a 
small cross-polarization component is contained in the electro-magnetic 
field and that the co- and cross-polarization are (slightly) phase shifted 
with respect to each other. In contrast, a perfectly linearly-polarized 
electro-magnetic field without cross-polarization content would result in 
single correlator values of '-1'. Compared to the histogram at 305~GHz, 
the histogram at 350~GHz shows a softened edge around the correlator value 
'0', extending into the positive-value domain of the histogram. 
This is due to the onset of the multimode propagation and the incipient
polarization rotation, leading to measured phase jumps in the detector current, 
as explained in Sec.~\ref{sec:052}. The experimental data corresponding to 
the lowest frequencies are not correctly described by the model. This is 
most likely due to the Gaussian beam profiles of the photo-mixer
which become non-ideal at these frequencies. In addition, we 
expect an influence from the vicinity of the propagation cut-off of the 
diagonal-horn antenna at about 180~GHz. The error bars quantify 
a small but measurable phase drift during the 
measurement.

\subsection{\label{sec:062}Higher-order modes}
Higher-order modes propagate in the waveguide from frequencies 
of $\approx 360$~GHz upwards. The total number of propagating modes is 
counted by the mode index $n(f)$ in Fig.~\ref{fig05}(a). 
Our multimode simulations excite at 
selected frequencies all possible, i.e.~energetically allowed, 
higher-order modes and through this we obtain, like before, sets of electric fields 
$(\boldsymbol{\mathcal{E}}_{\eta}, \boldsymbol{\mathcal{E}}_{\xi})$. 
We find in this case that the electric fields in the $\eta$- and 
$\xi$-direction are approximately of equal magnitude, 
cf.~Fig.~\ref{figS04}(c) and Appendix~\ref{app:D}. 
As a result of this, the polarization angle changes from $\beta \approx 15^\circ$ 
($\mathrm{TE}_{10}$ mode) to $\beta = 45^\circ$, cf.~Fig.~\ref{fig05}(c). 
The signature for this effect in the experiment are continuous correlator values 
between '1' and '-1', resulting in histograms of the type 
shown in Fig.~\ref{fig05}(a) at 400~GHz. Here, the histogram 
is characterized by balanced positive and negative values, consistent 
with the prediction of Sec.~\ref{sec:052}. Accordingly, the quotient 
of the sum over the positive and negative counts in the histogram, 
${\it pos/neg}\rightarrow 1$. We further find in far-field simulations 
that the phase difference between the $\eta$- and $\xi$-components 
of the electric fields of the same mode is negligible, 
cf.~Fig.~\ref{figS04}(d) of Appendix~\ref{app:D}, 
and similarly the phase differences between the electric-field components 
of different modes. Based on this, a simple 
multimode model can be established in which the quantity ${\it pos/neg}$ 
ratio directly relates to the polarization angle $\beta$. If ${\it pos/neg} = 1$, 
then $\beta = 45^\circ$ and if ${\it pos/neg}$ is smaller or larger than one, 
the polarization angle equals either $\arctan({pos/neg})$ or $\arctan({neg/pos})$. 
The latter indetermination of the polarization angle is due to the measurement 
procedure in which we rotate the waveguide by $90^\circ$, to 
measure the currents $I_{dc}^{E}$ and $I_{dc}^{H}$. Therefore, if 
${\it pos/neg}$ is not exactly equal to one, we cannot quantify whether 
the polarization direction was slightly larger or smaller than $45^\circ$. 
The blue data points in Fig.~\ref{fig05}(a) show the evaluation taking 
$\arctan({pos/neg})$. The other case is obtained by mirroring the 
blue data points with respect to $45^\circ$.
\section{\label{sec:07}Conclusion}
To conclude, we have shown that a diagonal-horn antenna, 
connected to a full-height rectangular waveguide, emits a linearly polarized 
electro-magnetic field, if the rectangular waveguide is excited by the 
$\mathrm{TE}_{10}$ mode. To confirm the field 
polarization experimentally we have used a 
method based on only a coherent detector and a 
planar silicon-mirror, acting as a Fresnel scatterer. This 
scheme is compatible with cryogenic experiments. 
At higher frequencies, we find that a multimode 
electro-magnetic field in the rectangular waveguide induces a polarization rotation 
by about $45^\circ$ of the emitted field from the diagonal-horn antenna, as 
confirmed by our simulations. The source of this polarization 
rotation is an advantageous mode topology in the rectangular waveguide.   
\begin{acknowledgements}
We acknowledge funding through the European 
Research Council Advanced Grant No.~339306 
(METIQUM). We like to thank Michael Schultz and the 
precision-machining workshop at the 
I.~Physikalisches Institut of the Universit{\"a}t zu K{\"o}ln 
for expert-assistance in the design and the fabrication 
of the diagonal-horn antennas and the waveguides. 
We also thank Anselm Deninger from 
TOPTICA Photonics AG, Germany, for extensive technical 
discussions. 
\end{acknowledgements}
\begin{appendix}
\section{\label{app:A}Measured detector response roll-off 
for a multimode sub-THz field}
Once the multimode sub-THz field is excited in the 
diagonal-horn antenna and is radiated from its aperture 
into free space, it consists of two fields, 
$\boldsymbol{\mathcal{E}}_{\parallel}$ and 
$\boldsymbol{\mathcal{E}}_{\perp}$, 
having equal magnitudes and polarizations perpendicular 
to each other. The resulting polarization direction is the 
vector sum of the polarizations of the two fields and is, 
hence, $45^\circ$ rotated compared to the polarization 
of the fundamental $\mathrm{TE}_{10}$ mode. Detecting 
the superposition of the two fields $\boldsymbol{\mathcal{E}}_{\parallel}$ 
and $\boldsymbol{\mathcal{E}}_{\perp}$, drives positive as well as 
negative currents in our coherent detector, that ideally 
exactly cancel each other. A signature of this effect would 
be a faster decrease of the detector current than one 
would expect due to its intrinsic roll-off. We measured 
the latter roll-off by a transmission 
measurement between source and detector only. 

In a first step, we have positioned the source and detector face-to-face and 
adjusted the distance between their apertures such 
that their beam waists lie on top of each other. By this 
we ensure maximum coupling between source and 
detector. 

\begin{figure}[tb!]
\centering
\includegraphics[width=\columnwidth]{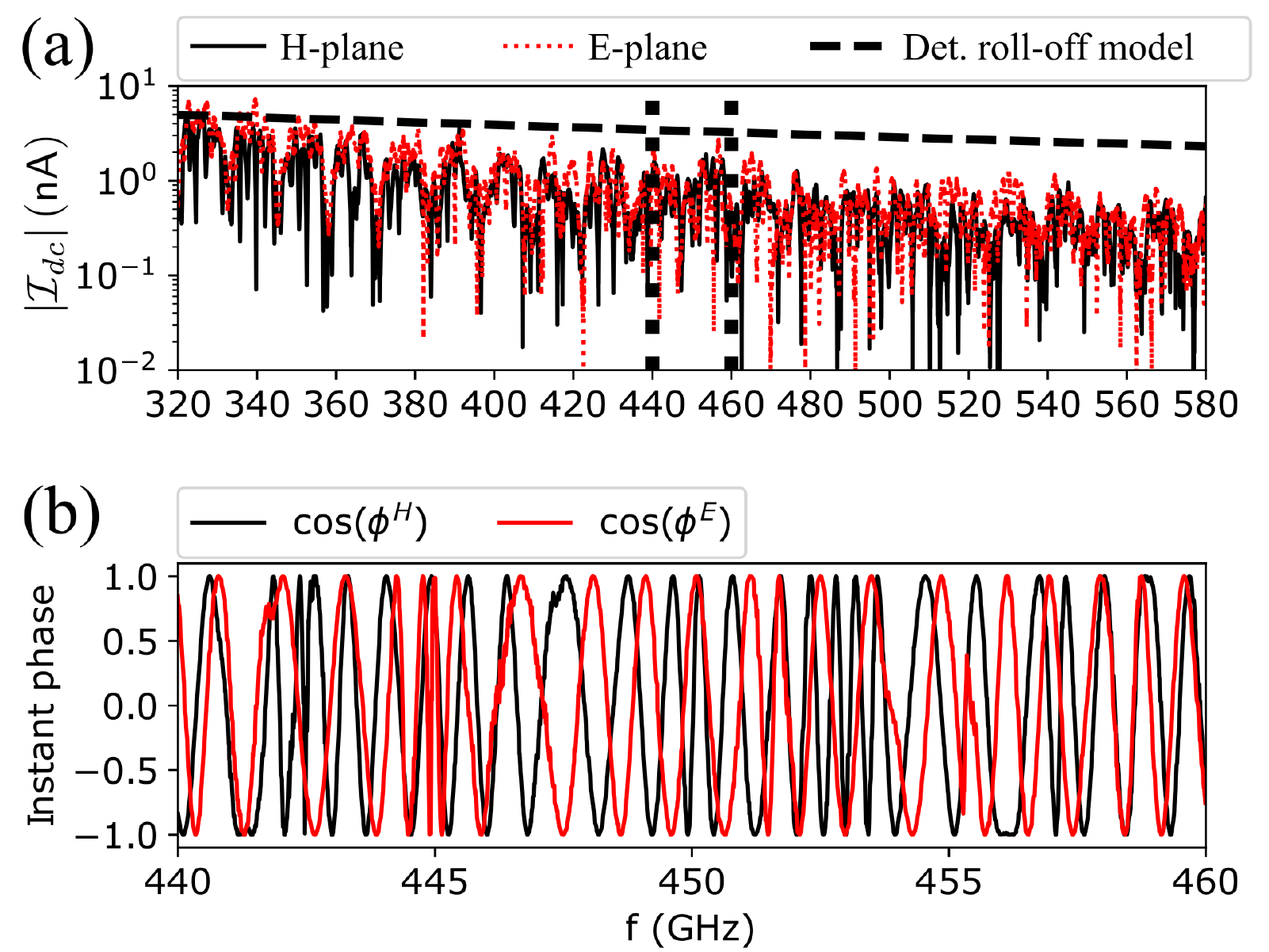}
\caption{\label{figS01}Detected multimode 
sub-THz field scattering off the $45^\circ$-tilted silicon mirror. 
(a) Measured detector response roll-off, 
expressed by $\lvert \mathcal{I}_{dc} \rvert$, Eq.~(\ref{eq:02}) of the main text. 
The black and the red trace show measurements when the H- or E-plane 
of the diagonal-horn antenna and waveguide are aligned parallel 
to the planar silicon-mirror plane of incidence. Note that the 
fluctuations of $\lvert \mathcal{I}_{dc}\rvert$ are due to the 
planar silicon-mirror transmission/reflection and partly 
also due to standing waves, but are not caused by the noise of 
the detector. The dashed line shows the measured Lorentzian 
detector response with roll-off time constant of 500~fs, determined 
by a transmission measurement between source and detector 
only. (b) Phase of the detected sub-THz field within a selected 
frequency range (dotted lines in (a)). Due to 
the detected multimode field, it shows phase jumps compared 
to the more ordered phase as a function of frequency in 
Fig.~\ref{figS02}(b) where only a single mode is detected. 
The data in the shown frequency window in (b) is employed to 
determine the datapoint at 450~GHz in Fig.~\ref{fig05}(a) of the main text.}
\end{figure}

In a second step, we have measured the detector 
response as a function of frequency between 150~GHz 
and 320~GHz. A Lorentzian curve of the form 
$A_{0}/\left[1+(2\pi f \tau)^2\right]$, with 
time constant $\tau = 500$~fs and $A_{0}$ a 
constant, fits the envelope of the detector response, 
$\tau$ being equal to the detector roll-off time. 
When comparing this intrinsic detector roll-off with 
the roll-off induced by the multimode field in the 
same detector, shown in Fig.~\ref{figS01}(a), we find 
that the latter decays faster. This is in-line with inducing 
positive as well as negative currents which, at least partly, 
tend to cancel each other. This effect leads primarily to phase jumps, 
shown in Fig.~\ref{figS01}(b), caused by 
suppressing the total detector current due to the 
multimode field.
\section{\label{app:B}Detection asymmetry $\mathcal{A}$ 
of parallel and perpendicular polarization components}
An ideal detector couples equally strong to the $\parallel$- and 
$\perp$-components of a received sub-THz field. In this case, one 
measures directly and only up to a coupling constant  
the Fresnel scattering amplitudes. 

\begin{figure}[tb!]
\centering
\includegraphics[width=\columnwidth]{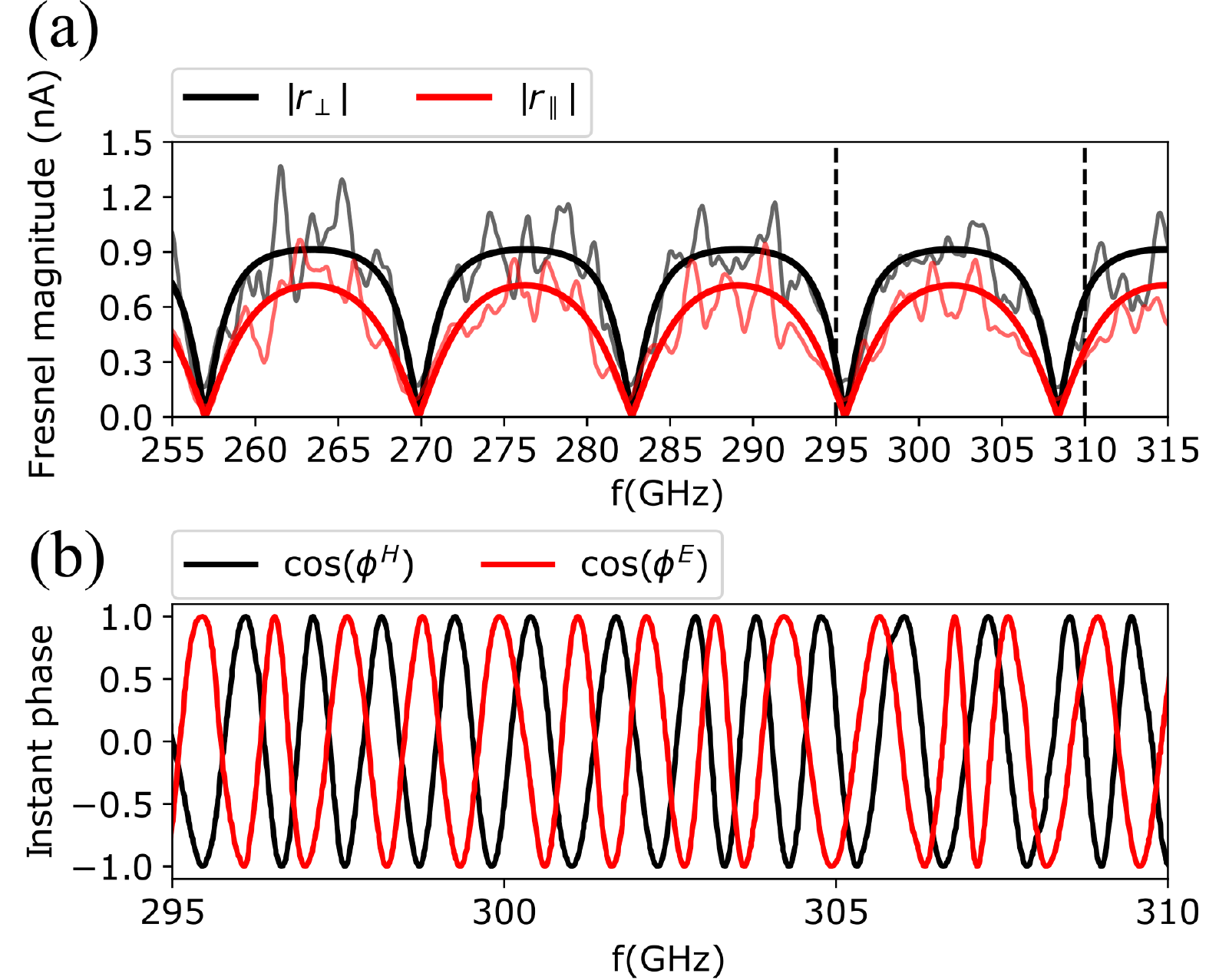}
\caption{\label{figS02}Detected single-mode 
sub-THz field scattering off the $45^\circ$-tilted silicon mirror. 
(a) Measured Fresnel magnitude 
of a detected free-space sub-THz field, being excited by 
the waveguide $\mathrm{TE}_{10}$ mode and radiated into 
free-space by the diagonal-horn antenna. 
(b) Measured instantaneous phase $\phi^{H,E}$, after scattering 
off the planar silicon-mirror. In (a) the smooth thick lines 
show the Fresnel theory and are compared with the experimental 
data (thin lines). We show the measured phase in (b) only 
within a selected frequency range (dashed region in (a)) 
for reasons of clarity.
With the E-plane of the waveguide and diagonal horn aligned 
parallel to the planar silicon-mirror plane of incidence (like shown 
in Fig.~\ref{fig04} of the main text), the scattering magnitude 
$\lvert r_{\parallel} \rvert$ and the phase $\phi^{E}$ are measured. 
Aligning the waveguide and diagonal horn H-plane parallel to the 
planar silicon-mirror plane of incidence, measures the scattering 
magnitude $\lvert r_{\perp} \rvert$ and 
the phase $\phi^{H}$. Note that in (a), the experimental 
data is rescaled to the magnitude of the theoretical 
Fresnel magnitudes which our experiment determines 
up to a constant factor. Furthermore, we added 
a factor $\mathcal{A} = 0.85$ to the experimental 
$\parallel$-component to match the theory as reasoned 
in Appendix~\ref{app:B}.}
\end{figure}

We evaluate the frequency dependence of these 
amplitudes, $r^{\parallel}(f)$ and $r^\perp(f)$,
using the Fresnel theory applied to the planar silicon-mirror. The 
mirror is characterized by the refractive index $n = 3.416$ and 
a thickness of $t = 3.415$~mm. For calibration purposes, we 
measured the amplitude reflection coefficients using the same 
coherent detector setup as described in the main text and confirmed their 
theoretically evaluated frequency dependence, 
cf.~Fig.~\ref{figS02}(a). Furthermore, we find also 
the expected Fabry-Perot resonance condition of the 
planar silicon-mirror which fully transmits the signal 
into the beam dump (element labeled 'BA' in 
Fig.~\ref{fig04} of the main text) at frequencies 
$pc_{0}/[2nt\cos\left(\alpha \pi/180\right)]$, 
with $p$ being an integer and $\alpha = 45^\circ$ is 
the angle of the planar silicon-mirror with respect to the axis of 
propagation of the input field. For these frequencies, both 
amplitude reflection coefficients are equal to zero.

We conducted this experiment and 
compared it to theory, in order to identify a possible asymmetry in the 
coupling to the $\parallel$- and $\perp$-components which we need to 
take into account in our modeling procedure. We find by this 
comparison that our detector couples to the $\parallel$-field 
component slightly stronger than to the $\perp$-field 
component. In order to compensate for this asymmetry, we need to 
multiply a factor $\mathcal{A} = 0.85$ to the experimentally determined
$\parallel$-component of the scattering amplitude to match it to the 
theoretical prediction.

In a second step we obtained the phases $\phi^{E}$ and $\phi^{H}$ 
after measuring the detector currents $I_{dc}^{E}$ and $I_{dc}^{H}$. 
Note the close to ideal phase shift of $\pi$ 
between the black and red trace in Fig.~\ref{figS02}(b) which show 
the cosine of the respective phases. This is also predicted by the 
Fresnel theory for the scattering of parallel and perpendicular 
polarizations off a dielectric layer. Deviations from the ideal phase 
shift condition are evident in our measurement as well and 
occur due to a number of reasons. First, a finite amount 
of cross-polarization in the detected beam, second, a relative phase 
shift (though being tiny) between co- and cross-polarizations and, third, the finite 
thickness $t = 3.415$~mm of the planar silicon-mirror which changes 
the relative phase shift between the reflected parallel and 
perpendicular components of the sub-THz wave, 
$\Delta \varphi^{\parallel - \perp}$, in Eqs.~(\ref{eq:03a}) 
and (\ref{eq:03b}) of the main text. 
\section{\label{app:C}Near- and far-field simulations - Part~I}
\begin{figure*}[tb]
\centering
\includegraphics[width=\textwidth]{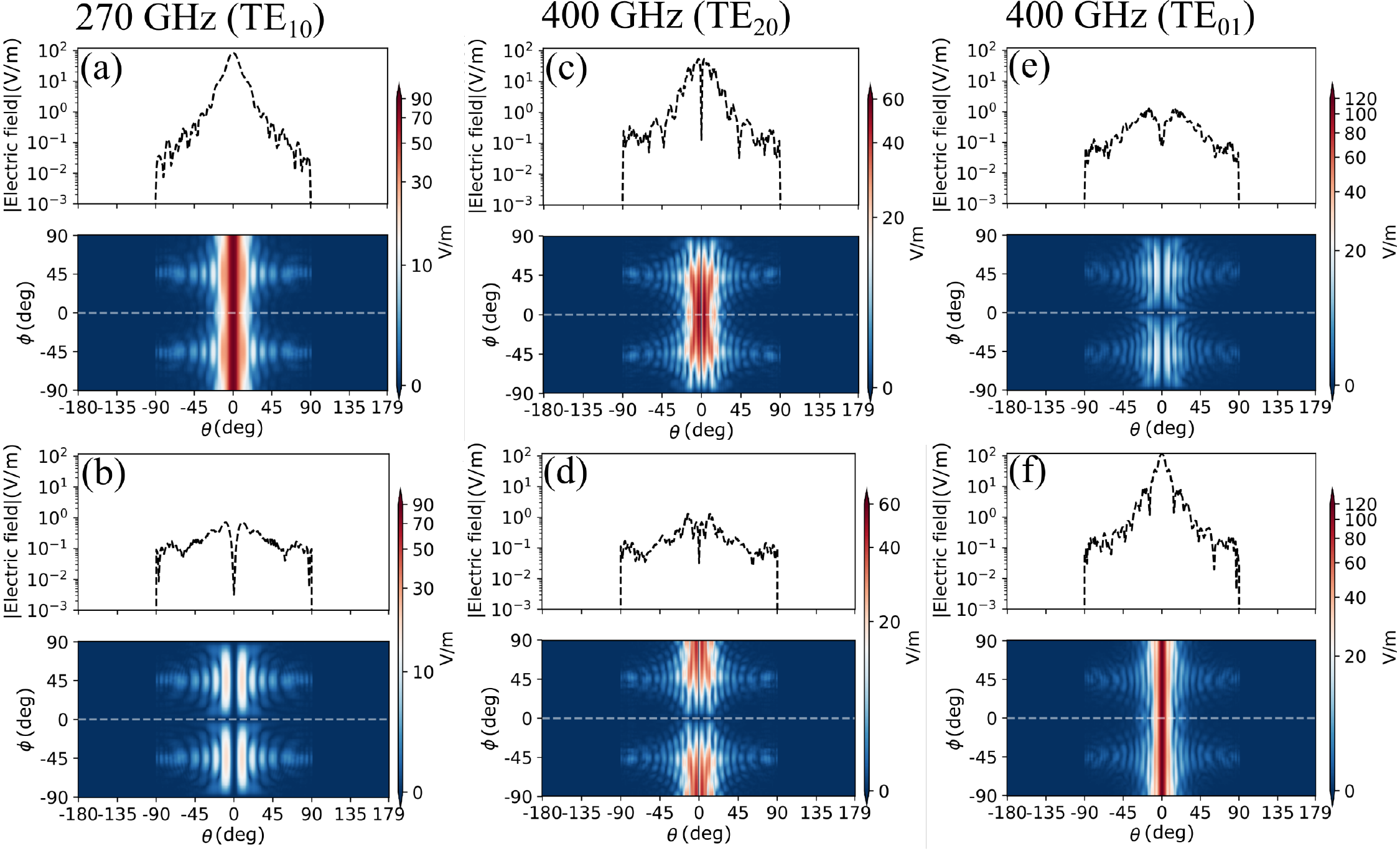}
\caption{\label{figS03}
Electro-magnetic far-field simulations 
of the diagonal-horn antenna at a 
reference distance of 1 meter from the aperture, 
performed with the software CST \cite{CST}. The 
images show the absolute value of the electric component 
of the far-field radiation pattern in the Ludwig-3 coordinate 
system \cite{ludwig1973}. The first three propagating 
modes emitted by the diagonal-horn antenna are shown at 
reasonably selected frequencies. The white dashed line indicates 
a cut through the map, shown on top of each figure. 
(a), (c) and (e) show the co-polarization component 
$\lvert\boldsymbol{\mathcal{E}}_{\eta}\rvert$ and (b), (d) and 
(f) show the cross-polarization component 
$\lvert\boldsymbol{\mathcal{E}}_{\xi}\rvert$, cf.~the definition of 
the $\eta$-$\xi$ aperture coordinate system in Fig.~\ref{fig01}(c) of the main text. 
The far-field intensities of the co- and cross-polarization 
components for the mode $\mathrm{TE_{10}}$ are 
approximately interchanged for the mode $\mathrm{TE_{01}}$ 
and the total absolute electric field in the co- and 
cross-polarization components for the mode $\mathrm{TE_{20}}$ 
is approximately the same. This indicates 
that for the superposition of the three shown propagating modes, the polarization 
is rotated by $45^\circ$ around the wavevector $k_{z}$.}
\end{figure*}

This section provides selected results of our electro-magnetic 
field simulations, modeling the diagonal-horn antenna output 
field. Figure~\ref{figS03} shows the far-field simulations 
of the diagonal-horn antenna, evaluating the radiated field 
by the antenna. The figures show by means of the field 
strengths in the co- and cross-polarization 
component (far-field) the mechanism of the {\it non-mechanical} 
polarization rotation, induced by the mode topology in the 
rectangular waveguide and in the diagonal-horn antenna. The figures 
focus only on the first three modes, in our scheme, the minimum number 
of modes necessary to induce the polarization rotation.
\section{\label{app:D}Near- and far-field simulations - Part~II}
\begin{figure}[tb]
\centering
\includegraphics[width=\columnwidth]{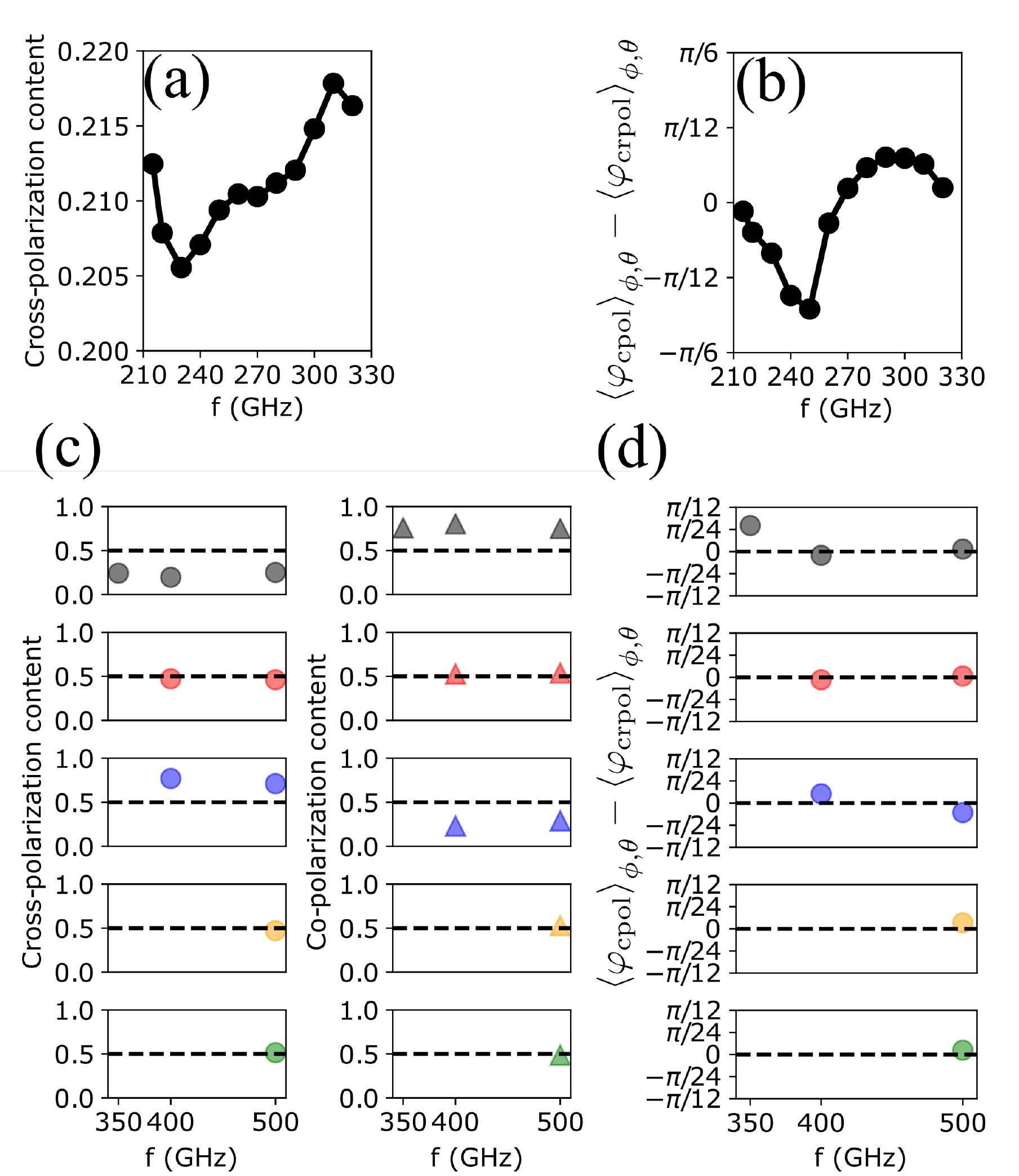}
\caption{\label{figS04}
Cross-polarization amplitude content 
$\mathcal{E}_{crpol}$, (a), and phase 
difference between the co- and cross-polarization 
($\langle \varphi_\mathrm{cpol} \rangle_{\phi,\theta}$ - 
$\langle \varphi_\mathrm{crpol} \rangle_{\phi,\theta}$) 
components, (b), for the 
$\mathrm{TE}_{10}$ mode. (c) Cross- and co-polarization 
amplitude content ($\mathcal{E}_{crpol}$ and 
$\mathcal{E}_{cpol}$) for the $\mathrm{TE}_{10}$ mode 
and higher order modes for selected frequencies 
(black $\widehat{=}~\mathrm{TE}_{10}$, red 
$\widehat{=}~\mathrm{TE}_{20}$, blue 
$\widehat{=}~\mathrm{TE}_{01}$, orange 
$\widehat{=}~\mathrm{TE}_{11}$ and green 
$\widehat{=}~\mathrm{TM}_{11}$). (d) Phase 
difference between the co- and cross-polarization 
components for the modes in (c), showing that the 
radiated polarization components in each mode are hardly phase 
shifted with respect to each other. The figure shows 
results only at selected frequencies, because of the 
numerically time-consuming simulations.
When the first three modes (black, red and blue) 
propagate for frequencies larger than 400~GHz, 
a nearly balanced amount of co- and cross-polarization 
amplitude components are indicative of a polarization rotation by 
$45^\circ$. The other two higher order modes, orange 
and green, contribute with equal parts to the co- and 
cross-polarization and, hence, further stabilize the 
polarization rotation by $45^\circ$.
}
\end{figure}

This section quantifies the co- and cross-polarization content in the 
calculated far-field radiation patterns. The final results are summarized in 
Fig.~\ref{figS04} and show the fundamental mechanism behind the 
polarization rotation we study in this paper. The data points in 
Figs.~\ref{figS04}(a) and (c) are obtained by integration of far-field 
patterns like shown in Fig.~\ref{figS03}. This obtains the co- and 
cross-polarization content, $A_{cpol}$ and $A_{crpol}$, 
of the electric component in the 
electro-magnetic field, radiated by the diagonal-horn antenna into the free-space. The data 
points in Figs.~\ref{figS04}(b) and (d) are obtained by $\phi, \theta$-integration 
in the Ludwig-3 coordinate system \cite{ludwig1973} 
of the calculated phase-front of the far-field.

Because the aforementioned calculation is important for predicting the 
polarization dynamics as a function of frequency, the remaining text provides 
an explanation of the employed formalism.

The field amplitudes in the polarization 
components are quantified by the integral of the respective 
absolute value of the sub-THz electric field 
over the polar coordinates $\theta$ and $\phi$ 
in the Ludwig-3 coordinate system \cite{silver_james1949, ludwig1973}, 
cf.~Fig.~\ref{figS03}. 
We express this formally as follows:
\begin{subequations}
\begin{align}
\label{eq:D1a}
A_{cpol} =\int_{S} \lvert{\boldsymbol{\mathcal{E}}_{\eta}}\rvert d\theta d\phi~,
\\
\label{eq:D1b}
A_{crpol} = \int_{S} \lvert{\boldsymbol{\mathcal{E}}_{\xi}}\rvert d\theta d\phi~,
\end{align}
\end{subequations}
where $\boldsymbol{\mathcal{E}}_{\eta}$ and $\boldsymbol{\mathcal{E}}_{\xi}$ 
are the electric far-fields, related to the $\xi$-$\eta$-aperture coordinate 
system of the diagonal horn, cf.~Fig.~\ref{fig01}(c), 
Fig.~\ref{fig03}(e)/(iv) and Fig.~\ref{fig05}(b), (c) of the main text. 
This coordinate system is introduced in order to 
avoid confusion with the $\parallel$- and $\perp$-components of 
the sub-THz field which are fixed space-wise (and with respect to the 
silicon-mirror plane of incidence) while the 
$\xi$-$\eta$ aperture coordinate system is fixed to the frame 
of the diagonal-horn aperture. For the fundamental 
$\mathrm{TE}_{10}$-mode, $\boldsymbol{\mathcal{E}}_{\eta}$ 
is largest and points in direction of the 
major polarization direction (co-polarization), 
whereas $\boldsymbol{\mathcal{E}}_{\xi}$ contributes 
to the much smaller cross-polarization. By rotating the 
diagonal-horn antenna one aligns either 
$\boldsymbol{\mathcal{E}}_{\eta}$ or 
$\boldsymbol{\mathcal{E}}_{\xi}$ with 
$\boldsymbol{\mathcal{E}}_{\parallel}$ or
$\boldsymbol{\mathcal{E}}_{\perp}$.

The integration area spanned by the polar coordinates is $S$ 
on which the far-fields in Fig.~\ref{figS04} are represented. The 
co- and cross-polarization content in the emitted sub-THz field 
is then determined by evaluating the 
expressions: 
\begin{subequations}
\begin{align}
\label{eq:D2a}
\mathcal{E}_{cpol} = \frac{A_{cpol}}{A_{cpol} + A_{crpol}}~,
\\
\label{eq:D2b}
\mathcal{E}_{crpol} = \frac{A_{crpol}}{A_{cpol} + A_{crpol}}~,
\end{align}
\end{subequations}
which are proportional to the field amplitudes in the 
E- and H-plane, cf.~Fig.~\ref{fig03}(e)/(iv) of the main text. The right-hand 
side of the equations are represented in Figs.~\ref{figS04}(a) 
and (c) for the different modes. Equations~(\ref{eq:D1a}), (\ref{eq:D1b}), 
(\ref{eq:D2a}) and (\ref{eq:D2b}) hold for a 
multimode field and quantify the ratio of 
$\boldsymbol{\mathcal{E}}_{\eta}$ and 
$\boldsymbol{\mathcal{E}}_{\xi}$ in the 
electro-magnetic field.

Higher-order propagating modes in the 
waveguide create a multimode (in our case 
up to five) electro-magnetic field. Interestingly, we have
discovered that this field has almost equal field strength 
in the components $\boldsymbol{\mathcal{E}}_{\eta}$ 
and $\boldsymbol{\mathcal{E}}_{\xi}$ and that these 
components are practically in-phase (cf.~Fig.~\ref{figS04}(d)), 
when radiated from the diagonal-horn antenna into 
free space. Furthermore, our multimode simulations 
find that the phase-delay between the aforementioned 
electric fields in different modes is practically negligible 
as well. This means, that 
the different modes are emitted by the diagonal-horn 
antenna in a coherent fashion and are practically not 
time-delayed with respect to each other. This 
is key for an effective polarization rotation of 
$45^\circ$ to happen and for generating a coherent 
electric field which is then radiated from the 
near- into the far-field by the diagonal-horn antenna. 
\section{\label{app:E}Imprinting Non-Classical States}
\begin{figure}[tb]
\centering
\includegraphics[width=\columnwidth]{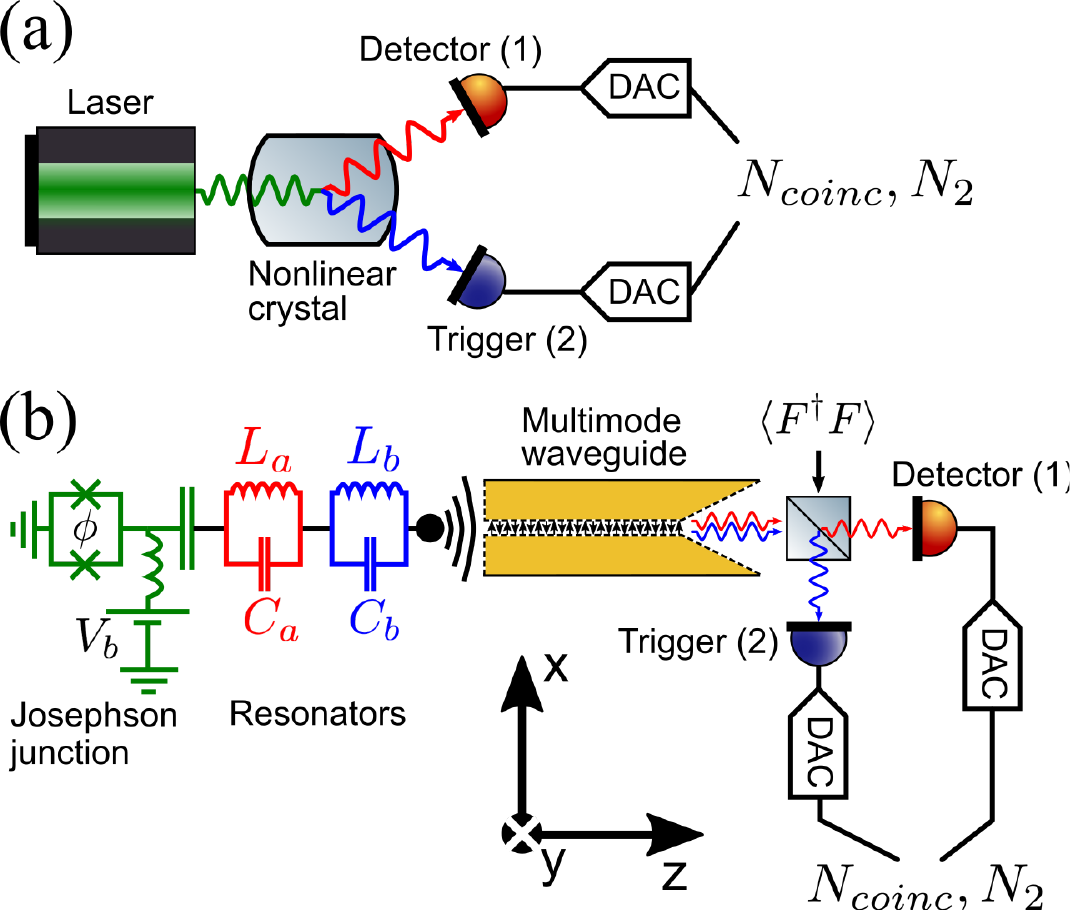}
\caption{\label{figS05}Analogy between photon-pair 
production during parametric down-conversion in a 
nonlinear crystal, (a), and due to charge-light coupling 
in a mesoscopic superconducting two-photon device, (b), 
like realized in 
\cite{westig2017, armour2015, leppakangas2013, trif2015} 
(green, red and blue parts of the figure). 
In (b), the red and blue photons excite the waveguide 
(for instance through chip-waveguide coupling \cite{westig2011}) 
and are radiated in the same or in different polarization states, 
dependent on their frequencies, into the 
detector and trigger apertures after scattering off a 
frequency-selective beamsplitter. 
The analog-to-digital converters (ADCs) count 
coincidences $N_{coinc}$ between detector and trigger 
and photon counts $N_{2}$ of the trigger alone. For 
detector characterization, the thermal photon population 
of the environment, $\langle F^\dagger F \rangle$, has 
to be considered.
}
\end{figure}

The presented results are of interest for single-photon detection 
using non-classical states of light, as discussed in the review 
by Ware et al.~\cite{ware2004}, which is of 
a very general nature and not directly 
tied to a specific frequency. The core idea presented in 
Ware et al.~\cite{ware2004} can be understood by means of 
Fig.~\ref{figS05}(a). Here, the central goal is to obtain the 
efficiency $\eta_{1}$ of the detector. It quantifies 
how efficient incoming single photons are recorded by the 
detector and, on the other hand, how many single photons are lost 
if the detector has not yet reached its fundamentally possible sensitivity.

A straightforward way is to make use of a 
nonlinear crystal, providing spontaneous parametric down-conversion 
(sPDC). As a nonlinearity one uses specific crystals that show a 
nonlinear polarization field response ('polarization' in this context 
means the polarization component of the electric displacement field 
and not the direction of the electric field studied in our paper) 
when strongly pumped by a laser. In order to exploit this effect, 
it is strongly {\it pumped} by a laser of frequency $f_p$ (green wave), 
and due to the nonlinear interaction, two photons (red and blue wave)
of different frequencies, named signal (s) and idler (i), are generated 
such that the energy and the momentum 
are conserved,
\begin{subequations}
\begin{align}
\label{eq:E1a}
2f_{p} &= f_{s} + f_{i}
\\
\label{eq:E1b}
\boldsymbol{k}_{p} &= \boldsymbol{k}_{s} + \boldsymbol{k}_{i}~.
\end{align}
\end{subequations}
The signal and idler photons are generated at the same time, they
are entangled and their power-correlation shows strong two-mode amplitude 
squeezing below the classical limit, hence, the name non-classical state.
Additionally, the outgoing wave polarization is ordinary or extraordinary 
with respect to the crystal axis.
One key feature of this non-classicality is that the detection of one photon 
out of a pair heralds the presence of the second one.
A second detector acting as a trigger, is used to 
record photon counts in coincidence with the detector, $N_{coinc}$, 
via two analog-to-digital converters (ADCs). Additionally, the 
ADCs measure also the photon counts $N_{2}$ of the 
trigger alone. Because signal and idler photons are generated at 
the same time, the remarkable advantage of this characterization 
technique is that the detector efficiency reads then 
simply \cite{ware2004},
\begin{equation}
\label{eq:E2}
\eta_{1} = \frac{N_{coinc}}{N_{2}}
\end{equation}
and is independent of the efficiency of the trigger.

However, sPDC using a nonlinear crystal is usually an inefficient 
process and a laser setup is needed. Moreover, due to the 
momentum conservation, Eq.~(\ref{eq:E1b}), the detector and trigger 
apertures have to be aligned correctly to receive all of the power in 
order to conduct a proper measurement. Also, the condition 
Eq.~(\ref{eq:E1b}) is frequency dependent such that the 
emission direction changes when the frequencies are adjusted. 
Additionally, the outgoing polarized fields usually do not have well defined 
Gaussian beam properties since the crystal influences the beam shape 
of the signal and idler fields.

A solution to these difficulties is provided by the 
two-mode non-classical source demonstrated by \citet{westig2017}.
It can be coupled by chip-waveguide coupling \cite{westig2011} to the 
waveguide in a setup proposed in Fig.~\ref{figS05}(b). Since the diagonal-horn 
antenna at the output of the multimode waveguide provides constant Gaussian beam
properties over a large frequency bandwidth, detector and trigger 
can be kept at constant position. Furthermore, in the example the two-mode non-classical 
source is based on the dynamical Coulomb blockade of a battery-powered 
Josephson junction coupled to a tailored electro-magnetic environment, 
therefore, complex laser setups are not needed.

Together with the progress reported in this 
paper on the polarization changes of the diagonal-horn antenna 
as a function of the frequency, a polarization sensitive detector can be 
characterized using the method of Ware et al.~\cite{ware2004}. 
Specifically, in our proposal a detector and a trigger 
would be employed which are only sensitive to linearly polarized 
electro-magnetic fields which is an often encountered technological situation. 
The task would be to measure the efficiency of such a
detector, only sensitive to a linear polarization. The emitted field of the 
diagonal-horn antenna, excited by the multimode waveguide, provides two 
options for such a measurement. First, when the signal (red) and the idler (blue) have 
frequencies such that only the $\mathrm{TE}_{10}$ mode is 
excited, detector and trigger have to be aligned in such a way to receive the 
same polarization direction. When the signal frequency remains in the 
$\mathrm{TE}_{10}$ mode but the idler frequency excites higher order modes, 
the trigger has to be rotated by $45^\circ$ with respect to the detector, 
cf.~Fig.~\ref{fig05}(b) and (c). The separation of the different frequencies is 
achieved by a frequency selective beamsplitter. At the open port of the 
beamsplitter a thermal photon population is important to quantify, when 
only a single detector setup would be used for characterization of the 
detector efficiency. For a correlation setup measuring coincidences like 
proposed by Fig.~\ref{figS05}(b), the thermal photon population does not 
influence the measurement outcome since it is not correlated at two 
different frequencies.
\end{appendix}
\end{document}